\documentclass[12pt]{article}
\usepackage{algorithm, algpseudocode}
\usepackage{amssymb, amsmath, amsthm}
\usepackage{mathabx}

\usepackage{graphicx}
\usepackage{subcaption}
\usepackage{tikz}
\usepackage{float}
\usepackage[margin=1in]{geometry}
\usepackage{microtype}
\usepackage{multirow}

\newcommand{\argmin}{\operatornamewithlimits{arg\,min}}

\newcommand{\indep}{\perp\!\!\!\perp}
\date{}

\usepackage{mathtools}

\renewcommand{\tilde}{\widetilde}
\renewcommand{\hat}{\widehat}

\usepackage[colorlinks,citecolor=blue,urlcolor=blue,linkcolor=red,naturalnames=true]{hyperref}
\usepackage{natbib}

\usepackage{amsmath}
\theoremstyle{plain}

\newtheorem{prop}{Proposition}

\usepackage{setspace}
\usepackage{blindtext}
\usepackage{graphicx} 
\usepackage{tocloft}
\usepackage{xr}

\title{\textbf{Constrained convex clustering for interpretable spatial domain detection in spot-based spatial transcriptomics}}
\author{Hyun Jung Koo\footnote{koo00029@umn.edu} \ and Aaron J. Molstad\footnote{corresponding author: amolstad@umn.edu}\\
School of Statistics, University of Minnesota\bigskip\\
\date{}}

\begin{document}

\maketitle
\vspace{-40pt}
\begin{center}
  {\small
    \textbf{Funding.} A. J. Molstad's contributions were supported by a grant from the National Science Foundation (DMS-2413294). H. J. Koo's contributions were supported by the the DSI-MnDRIVE Graduate Assistantship Award at the University of Minnesota. \\
    \textbf{Conflict of interest.} The authors declare no competing interests.
  }
\end{center}
\begin{abstract}
Popular technologies for generating spatially resolved transcriptomic data measure gene expression at the resolution of a ``spot'', i.e., a small tissue region 55 $\mu$m in diameter. Each spot can contain many cells of different types. In typical analyses, researchers are interested in using these data to identify and profile discrete spatial domains in tissue. In this paper, we propose a new method, DUET, which simultaneously identifies discrete spatial domains and estimates each spot's expected cell-type proportion. This allows the identified spatial domains to be characterized in terms of the underlying expected cell-type proportions, which affords interpretability and biological insight. DUET utilizes a constrained version of model-based convex clustering, and as such, can accommodate Poisson, negative binomial, normal, and other types of expression data. Moreover, our convex clustering-type criterion allows for both the number of clusters and degree of spatial smoothness to be controlled by a single tuning parameter, which can be chosen in a data-driven fashion. Through simulation studies and a real data application, we show that DUET can achieve better clustering and deconvolution performance than some existing methods.
\end{abstract}

\setstretch{1.3}
\section{Introduction}
\label{s:intro}

Recent advances in spatially resolved transcriptomics (SRT) have transformed our ability to study gene expression within intact tissue architecture, providing insight into the complex relationship between cellular heterogeneity and spatial organization \citep{staahl2016visualization}. Spatial transcriptomic profiling combines high-throughput sequencing with spatial barcoding, where RNA molecules are tagged with positional codes during cDNA synthesis, allowing gene expression to be mapped back onto specific two-dimensional coordinates on the tissue \citep{moses2022museum}. Technologies such as 10x Visium capture transcriptome-wide data while preserving spatial context, enabling comprehensive profiling of tissue microenvironments \citep{10xGenomics2020}. The Visium workflow involves (i) tissue fixation and sectioning (typically 5-10 $\mu$m thick), (ii) permeabilization to release RNA onto barcoded spots, (iii) reverse transcription with spatial barcodes, and (iv) library preparation for sequencing \citep{williams2022introduction}. Each spot covers up to 50 cells, depending on tissue density. Spots are arranged on a grid with 100 $\mu$m center-to-center spacing, which trades off spatial resolution against transcript capture efficiency \citep{saunders2018molecular}.
However, the heterogeneity of the cells within a spot poses a fundamental analytical challenge: the measured gene expression is an aggregate over all cells in a spot, and those cells may be of different, unknown types.

The challenge of resolving mixed transcriptional signals is not unique to spatial sequencing technologies. Before the rise of spatial transcriptomics, bulk RNA-seq deconvolution methods like CIBERSORT \citep{newman2015robust} were developed to estimate cell-type proportions from mixed transcriptional profiles. These approaches leverage reference profiles, increasingly from single-cell RNA sequencing (scRNA-seq) data \citep{baron2016single}, to computationally decompose bulk measurements into estimated cellular abundances \citep{avila2018computational}. Spatial deconvolution extends this paradigm by associating estimated cell-type proportions with specific tissue locations.

To address this challenge, numerous spatial deconvolution methods have been developed that leverage scRNA-seq as a reference. For example, SPOTlight \citep{elosua2021spotlight} and Stereoscope \citep{andersson2020single} estimate cell-type proportions by modeling spot-level gene expression as a mixture of scRNA-seq-derived signatures. More recently, \citet{ma2022spatially} introduced CARD, which uses a conditional autoregressive prior to share information across neighboring spots.

In parallel, spatial clustering methods such as Giotto \citep{dries2021giotto} and BayesSpace \citep{zhao2020bayesspace} have been developed to identify spatially coherent domains based on expression patterns. However, current analytical paradigms typically treat deconvolution and clustering as sequential and separate steps. CARD, for example, performs only deconvolution, using spatial smoothing to improve spot-level estimates; clustering methods like Giotto perform only domain detection, operating either on raw expression or on post hoc–aggregated deconvolution outputs. This decoupled approach can limit the biological interpretability of the identified spatial domains, since it does not jointly account for transcriptional similarity and cellular composition. Moreover, most existing frameworks do not provide formal estimates of cell-type distributions at the cluster level. Of the methods that do perform simultaneous deconvolution and clustering, e.g., IRIS \citep{ma2024accurate}, there are key methodological differences from our proposed method, which we discuss in Section \ref{s:comparison_methods}.  

To address these limitations, we propose a novel model-based framework for joint deconvolution and discrete domain detection with SRT data. Our approach is built on a new version of convex clustering \citep{hocking2011clusterpath,chi2015splitting,wang2021integrative,tan2015statistical,sun2021convex,chi2025and} that accounts for both the spatial organization of spots and the compositionality of the parameters. This lets us estimate expected spot-level cell-type compositions and identify biologically interpretable spatial domains in a single step.  
Through simulation studies and real data analyses, we demonstrate that our method consistently outperforms sequential pipelines and existing methods, especially in settings with fine-grained tissue structure or subtle spatial variation. 

\vspace{-10pt}
\section{Methodology}
\label{s:methods}
\subsection{Model assumptions}
In this paper, we treat measured expression for the $g$th gene in the $i$th spot, $x_{gi},$ as a realization of a random variable whose distribution can be described in two parts. In particular, given the \textit{realized} cell-type composition for the $i$th spot, $v_i$, we assume the conditional expectation
\begin{equation}\label{mod:hierarchical}
{\rm E}(X_{gi} \mid V_{i} = v_i) = s_i^* \cdot b_g^\top v_i ~~~ \text{ with }~~~{\rm E}(V_i) = \theta^*_i,
\end{equation}  
where $b_g \in \mathbb{R}^K_+$ is the average expression of gene $g$ in the $K$ different cell types,  $\theta^*_i \in \Delta^{K - 1}$ is the \textit{expected} cell-type composition for the $i$th spot, $s^*_i \in \mathbb{R}_+$ is a size factor for the $i$th spot, and $v_i$ is  the realized cell-type composition for the $i$th spot. Here, the cell-type composition vectors $\theta^*_i$ and $v_i$ belong to the set $\Delta^{K-1} = \{v \in \mathbb{R}^K: v_j \geq 0, 1_K^\top v = 1\}$ and $\mathbb{R}_+$ denotes the set of nonnegative real numbers. Naturally, ${\rm E}(X_{gi}) = {\rm E}\{{\rm E}(X_{gi} \mid V_{i})\}= s_i^* \cdot b_g^\top \theta^*_i$. 

In this work, we define a spatial domain as a set of spots that have identical expected cell-type composition vectors. That is, the $c$th spatial domain is the set of spots $\{i \in [n]: \theta^*_i = \theta^\ddagger_{c}\}$ for distinct compositions $\theta^\ddagger_{1}, \dots, \theta^\ddagger_{C}$. The goal of our method is thus to simultaneously estimate $\theta^*_1, \dots, \theta^*_n$ and discover the spatial domains through their characteristic compositions $\{\theta^\ddagger_{c}\}_{c=1}^C$. To be clear, the goal of our method is not to estimate each spot's realized cell-type composition, but rather to provide a parsimonious representation of the expected cell-type composition across the entire tissue. We view this task as more biologically insightful, as it targets the underlying tissue architecture that governs cell-type composition across the tissue, rather than its realization at any single spot.

Our method assumes that $X_{gi} \indep X_{gi'}$ for all $i \neq i'$. This may seem counterintuitive since in practice, nearby spots tend to
have similar expression. In our model, however, spatial similarities arise from similarities in spot-specific expected cell-type compositions $\Theta^* = (\theta^*_1, \dots, \theta^*_n)^\top$. 
We make this assumption in part to obtain a tractable loss function for $\theta^*_i$, as we discuss later in this section. The resulting estimating equations for $\Theta^*$ remain unbiased (even if independence does not hold) as long as the first moment ${\rm E}(X_{gi}) = s^*_i \cdot b_g^\top \theta^*_i$ is correctly specified.

Consistent with common modeling assumptions for RNA-seq data, we will focus on three specific distributions for the $X_{gi} \mid V_i$: normal, Poisson, and negative binomial. The data $X_{gi}\mid V_i$ can be either the count data from the RNA-seq experiment, or normalized variations thereof.
In each case, we will assume that ${\rm E}(X_{gi} \mid V_i = v_i) = s^*_i \cdot b_g^\top v_i \in \mathbb{R}_+$. 
As discussed in the introduction, the reference matrix $B$ is typically estimated from available scRNA-seq reference data (see Supplementary Materials Section 1 for details). Consequently, the size factor $s^*_i$ can roughly be interpreted as the number of cells in the $i$th spot. Finally, suppose that we have restricted our attention to genes that are independent so that $X_{gi} \indep X_{g'i}$ for $g \neq g'$ for all $g \in [G]$ and $i \in [n]$. 

To gain intuition for our estimation approach, consider the setting where $X_{gi} \mid V_i$ is normal with variance $\sigma^*_{gi}$ known, in which case the maximum likelihood estimator of $(s^*_i, v_i)$ would be
$\argmin_{s_i \geq 0, v_i \in \Delta^{K-1}} \sum_{g=1}^G (x_{gi} - s_i \cdot b_g^\top v_i)^2/(2\sigma_{gi}^{*2})$. 
We solve this through reparameterization by taking $\rho^*_i = s^*_i \cdot v_i$, which converts the constrained  
  problem to the nonnegative least squares problem
\vspace{-5pt}
\begin{equation}\label{eq:spot-by-spot}
     (\widebar{s}_i, \widebar{v}_i)  = \left(\widebar{\rho}_{i}^\top 1_K, \frac{\widebar{\rho}_i}{\widebar{\rho}_{i}^\top 1_K}\right), ~~~ \text{ where }~~~\widebar\rho_i \in \argmin_{\rho_{i} \geq 0} \sum_{g=1}^G \frac{(x_{gi} - b_g^\top \rho_i)^2}{2\sigma_{gi}^{*2}},
\end{equation}
assuming that $\widebar\rho_i$ is not the zero vector. If each $\theta_i^*$ were distinct, then $\widebar{v}_i \equiv \widebar{\theta}_i$ would be a natural estimator of $\theta_i^*$, since $\text{E}(X_{gi}) = s_i^* \cdot b_g^\top \theta_i^*$ implies that the conditional score is unbiased at $(s_i^*, \theta_i^*)$ by iterated expectation. That is, if $\ell_i(v_i, s_i)$ is the $i$th spot's likelihood contribution corresponding to the $X_{gi} \mid V_i = v_i$, then ${\rm E}\left[\nabla_{v_i} \ell_i(\theta^*_i, s^*_i)\right] = 0$ for each $i$, where the expectation is taken with respect to the marginal of $X_{i}$.
However, estimating each $\theta_i^*$ separately in this way ignores spatial structure: if nearby spots share the same expected composition, this should be leveraged to improve efficiency. 

If some of the nearby $\theta_i^*$ are identical, maximum likelihood estimation would require specifying the full distribution of each $V_i$, not just its mean, and would knowledge of which $\theta_i^*$ are identical apriori. To avoid this, we adopt the conditional likelihood of $X_{i} \mid V_i$ as a loss function for $\theta_i^*$. As mentioned, this is justified by the fact that, under our marginal moment specification $\mathrm{E}(X_{gi}) = s_i^* \cdot b_g^\top \theta_i^*$, the conditional score is an unbiased estimating equation for $\theta_i^*$: by iterated expectation, its expectation vanishes at $(\theta_i^*, s_i^*)$ whenever $\mathrm{E}(V_i) = \theta_i^*$, without further distributional assumptions on $V_i$. This avoids committing to a model for $V_i$, making spatial domain recovery robust to misspecification of the compositional heterogeneity across spots. We verify in simulation (Section \ref{sec:Simulations}) that this choice yields accurate recovery of both $\theta_i^*$ and, perhaps unsurprisingly, the realized $v_i$.

The objective of our proposed method is to estimate the cell-type composition matrix $\Theta^*$ in a way such that we can simultaneously leverage the spatial information to substantially improve efficiency relative to spotwise estimation, and perform spatial domain detection of the spots according to their cell-type composition (i.e., hard clustering of the $\hat{\theta}_i$).

\vspace{-10pt}
\subsection{Penalized M-estimation criterion}
To achieve simultaneous spatial domain detection and cell-type deconvolution, we use a penalized M-estimation criterion. Building on the spot-by-spot deconvolution estimation criterion described in the previous section, we introduce a convex penalty that can encourage the estimates of $\theta^*_i$ to be exactly identical for nearby spots. Let $\mathcal{F}_i(\theta, s) = -  \log f(x_i; \theta, s)$ where $f$ is the density corresponding to $X_i = (X_{1i}, \dots, X_{Gi})^\top$ given $V_i = \theta$ and $s$ at $x_i$. The estimator of $\Theta^*$ and $S^* = (s_1^*, \dots, s_n^*)^\top$ we propose is $(\widehat{\Theta}^\lambda, \widehat{S}^\lambda)$, defined as
\begin{equation}\label{eq:general_estimator0}
    \argmin_{\Theta \in \mathbb{R}^{n \times K}, S \in \mathbb{R}^n} \left\{\frac{1}{n} \sum_{i=1}^n \mathcal{F}_i(\theta_i,s_i) + \lambda \sum_{i < j} \gamma_{ij} \|\theta_i - \theta_j\|_2\right\}~~ \text{ subject to }~ \theta_i \in \Delta^{K-1}, s_i \geq 0 ~~ \forall  i \in [n],\end{equation}
where $\lambda \geq 0$ is a tuning parameter, and $\gamma_{ij} \geq 0$ is a weight that accounts for distance between spots $i$ and $j$. The group-lasso-like penalty in \eqref{eq:general_estimator0} is nondifferentiable when $\theta_i = \theta_j$, so when $\lambda$ and $\gamma_{ij}$ are sufficiently large, the solution to \eqref{eq:general_estimator0}, say $\widehat{\Theta}^\lambda$, will be such that $\widehat{\theta}_i^\lambda = \widehat{\theta}_j^\lambda$.  Through the exact fusion property of the Euclidean norm penalty, our estimator performs spatial domain detection and cell-type estimation simultaneously. We call \eqref{eq:general_estimator0} DUET (or D$^3$UET) for \underline{d}omain \underline{d}etection and \underline{d}econvolution \underline{u}nified \underline{e}s\underline{t}imator.

\begin{figure}[t]
    \centering
    \includegraphics[width=\textwidth]{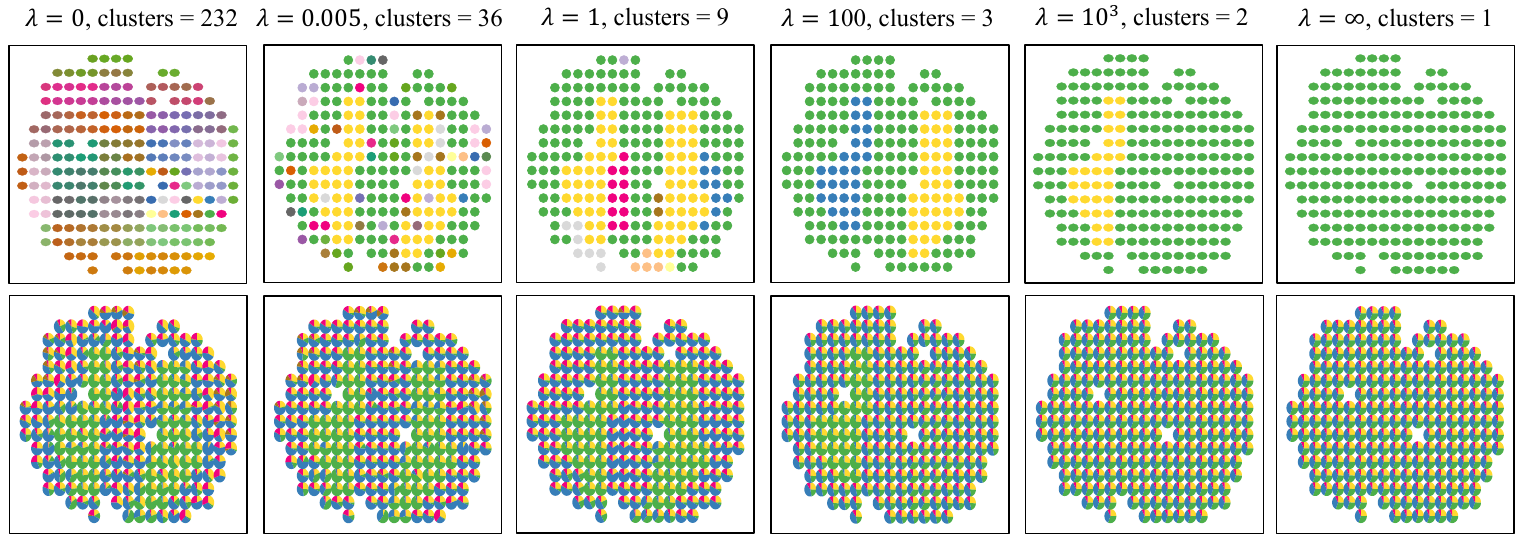}
    \caption{Illustration of DUET clustering path on mouse olfactory bulb data. In the top row, each distinct color represents a distinct cluster, and all spots from the same cluster have identical estimates of their expected cell-type proportion. As the tuning parameter $\lambda$ increases from 0 to $\infty$, the number of clusters decreases from $n$ to 1. In the bottom row, we show the corresponding estimated expected cell-type proportions for each spot.}
    \label{fig:tuning_path}
\end{figure}

The regularization technique used in \eqref{eq:general_estimator0} is essentially a generalization of convex clustering \citep{chi2025and}. 
Compared to traditional clustering methods (e.g., $k$-means), convex clustering offers several key advantages. First, it does not require pre-specifying the number of clusters since the fusion of spots occurs naturally as the tuning parameter $\lambda$ increases, producing a continuous solution path from $n$ distinct clusters (when $\lambda=0$ so each $\hat\theta_i^0 = \widebar{\theta}_i = \widebar{v}_i$) to a single cluster (when $\lambda$ is sufficiently large) as seen in Figure \ref{fig:tuning_path}. This avoids the need to subjectively pre-specify the number of clusters: we need only select $\lambda$, which we do in an objective and data-driven way. 
Second, unlike many clustering approaches currently used for spatial transcriptomics (e.g., Seurat) that operate on dimension-reduced representations of the data (e.g., principal components), convex clustering works directly with the raw data, preserving the interpretability and avoiding loss of information.

In contrast with existing generalizations of convex clustering \citep{wang2021integrative}, \eqref{eq:general_estimator0} is distinct in that the clustered parameters $\theta_i$ are constrained to the unit simplex, and they enter the nfegative log-likelihood through their product with the $i$th spot's size factor $s_i$. The size factors themselves are unpenalized and do not appear in the penalty term. If, for example, $\lambda = 0$, then the solution is simply the spot-by-spot maximum likelihood deconvolution estimator.

We specify the weights $\gamma_{ij}$ in a way such that only adjacent spots contribute to the penalty in \eqref{eq:general_estimator0}. We provide more details in Section \ref{sec:weight_construction}, and illustrations of the weights in Section \ref{sec:PDAC}.

To make \eqref{eq:general_estimator0} concrete, for the remainder of the article we will focus on our estimator's application to the Poisson model with identity link. In this case, our estimator is 
\begin{equation} \label{eq:general_estimator}
    (\widehat{\Theta}^\lambda, \widehat{S}^\lambda) \in \argmin_{\Theta \in \mathbb{R}^{n \times K}, S \in \mathbb{R}^n} \left[\frac{1}{n} \sum_{i=1}^n \sum_{g=1}^G \{s_i \cdot b_g^\top \theta_i - x_{gi}\log(s_i \cdot b_g^\top \theta_i)\} + \lambda \sum_{i < j} \gamma_{ij} \|\theta_i - \theta_j\|_2\right]
\end{equation}
$$ \text{ subject to }~~ \theta_i \in \Delta^{K-1}~~ \text{and} ~~ s_i \geq 0 ~~~~\text{for all } i \in [n].$$
As written, this model requires that the entries of $B$ are positive, which is common in practice. If any entries of $B$ are zero, we found that replacing them with small pseudocounts (e.g., $10^{-6}$) has a negligible effect on our method across a variety of settings. Alternatively, we could replace the mean function $s_i \cdot b_g^\top \theta_i$ with $s_i \cdot b_g^\top \theta_i + \epsilon$ for small positive $\epsilon$.  

\subsection{Related methods}
\label{s:comparison_methods}
A closely related method is IRIS \citep{ma2024accurate}, which also performs simultaneous cell-type deconvolution and spatial domain detection, though it focuses primarily on recovering spatial domains and the realized cell-type proportions. Letting $\tilde{X} \in \mathbb{R}^{G \times n}$ be a size-factor standardized matrix of gene expression and letting $\mathcal{D}$ denote the set of pairs of nearby spots (see Supplementary Materials Section 4.1), the IRIS estimator is defined as
\begin{equation}\label{eq:IRIS_criterion}
    \argmin_{V, Q, \mu}\left\{\|\tilde{X} - BV^\top\|_F^2 + \lambda_1 \sum_{r=1}^R \|V^\top (Q_r/1^\top Q_r) - \mu_r\|_2 + \lambda_2 \sum_{(i,j) \in \mathcal{D}} \|v_i - v_j\|_2^2\right\},
\end{equation}
where $R, \lambda_1$, and $\lambda_2$ are tuning parameters with $R$ being the number of distinct spatial domains. In \eqref{eq:IRIS_criterion}, the optimization variables are $V$, the matrix of realized (unobservable) cell-type proportions for each spot; $\mu = (\mu_1, \dots, \mu_R)$, the centroid vectors of each of the $R$ distinct spatial domains; and $Q = (Q_1, \dots, Q_R) \in \mathbb{R}^{n \times R}$ is a matrix consisting of binary vectors indicating whether each spot belongs to spatial domain $r$. DUET is not a special case of IRIS because IRIS uses a differentiable ridge-type penalty on the $v_i$, and a discontinuous penalty on the $\mu_r$. Moreover, IRIS does not impose simplex constraints during optimization; nonnegativity and sum-to-one are enforced as a post-processing step. On the whole, we view IRIS and DUET as fundamentally distinct methods for spatial domain detection: IRIS targets the realized cell-type proportions, while DUET targets the expected cell-type proportions. 

DUET and IRIS also differ in how they are tuned in practice. IRIS involves three tuning parameters: $\lambda_1$, $\lambda_2$, and the number of spatial domains $R$. In practice, the publicly available IRIS software exposes only $R$ to the user, with $\lambda_1$ and $\lambda_2$ set to internal defaults. However, $\lambda_1$ and $\lambda_2$ can have a substantial impact on model fit (see Supplementary Materials Section 4.2 for a sensitivity analysis). In DUET, the number of spatial domains and the degree of spatial smoothness are jointly governed by a single parameter $\lambda$, which we select in a data-driven manner via an information criterion.

Several other methods exist for spatial domain detection and cell-type deconvolution, including GraphST \citep{long2023spatially}, STitch3D \citep{wang2023construction}, and SPACEL \citep{xu2023spacel}. These approaches rely on graph neural networks or deep embeddings to learn spatial representations, and generally do not provide interpretable cell-type composition estimates at the domain level. Because DUET's focus is on model-based, interpretable joint estimation of expected compositions and domains, we restrict our comparisons to methods that share this goal (IRIS, CARD, BayesSpace) or are standard pipelines in practice (Seurat, SPOTlight).

\section{Computation}
\label{s:computation}
\subsection{Overview}
In this section, we discuss the algorithm for computing \eqref{eq:general_estimator}, though applications to other data generating models are straightforward. The DUET optimization problem is nonconvex in general, but biconvex: it is convex in $s_i$ with $\theta_i$ held fixed, and convex in $\theta_i$ with $s_i$ held fixed.
Therefore, we use a two-block coordinate descent algorithm to solve this biconvex problem. The update for each component of $S = (s_1, \dots, s_n)^\top \in \mathbb{R}^n_+$ is equivalent to solving a nonnegative generalized linear model estimation problem, and the update for $\Theta$ is a constrained version of generalized convex clustering. 
In a subsequent section, we will describe the update for $\Theta$ with $S$ held fixed. The general structure of the two-block coordinate descent algorithm is summarized in Algorithm \ref{alg:DUET_Combined}.

Note that 
with \( \Theta \) fixed, the update for \( S \) can be separated across the entries \( s_i \). In particular, the update for \( s_i \) is given by
$
s_i^+ = \argmin_{s_i \geq 0}~ \sum_{g=1}^G \left\{s_i \cdot b_g^\top \theta_i  - x_{gi} \log \left( s_i \cdot b_g^\top \theta_i  \right) \right\}$. Setting the gradient to zero gives the closed-form update
$
s^+_i = \sum_{g=1}^G x_{gi}/\sum_{g=1}^G b_g^\top \theta_i, i \in [n],
$
which will be nonnegative when  $x_{g'i} > 0$ and $ b_g^\top \theta_i > 0$ for some $g$ and $g'$. The former can be enforced by only including genes with some nonzero counts; the latter is enforced naturally when $B$ is nonnegative and $\Theta$ is initialized in the feasible set. 

\begin{algorithm}[t]
\caption{Two-block coordinate descent with proximal ADMM for computing DUET}
\label{alg:DUET_Combined}
\begin{algorithmic}[1]
\small
\Require $\mathbf{X} \in \mathbb{R}_+^{G \times n}$ (SRT data), $\mathbf{B} \in \mathbb{R}_+^{G \times K}$ (from scRNA-seq reference)
\Require Initial nonnegative matrices $\Theta^{(0)} \in \mathbb{R}_+^{n \times K}, S^{(0)} \in \mathbb{R}_+^{n}$
\Require Tolerances $\epsilon_{\text{rel}} > 0, \epsilon_{\text{pri}} > 0, \epsilon_{\text{dual}} > 0$
\State $k \gets 0$
\Repeat
    \For{$i \in [n]$ in parallel}
    \State $s_i^{(k+1)} \gets \argmin_{s_i \geq 0}~ \mathcal{F}_i(\theta_i^{(k)}, s_i)$
    \EndFor
    
    \State \textbf{Initialize ADMM:} $r \gets 0$, set $\Theta^{(k, 0)} \gets \Theta^{(k)}$, and initialize $\Omega^{(k, 0)}, \Gamma^{(k, 0)}$
    \While{not converged}
        \State $\Theta^{(k, r+1)} \leftarrow \argmin_{\Theta} \mathcal{L}_\rho(\Theta, \Omega^{(k, r)}, \Gamma^{(k, r)}) + \frac{\rho}{2}{\rm tr}\left[(\Theta - \Theta^{(k, r)})^\top (\eta I_n - A^\top A)(\Theta - \Theta^{(k, r)})\right]$ 
        \State $\Omega^{(k, r+1)} \leftarrow \argmin_{\Omega} \mathcal{L}_\rho(\Theta^{(k, r+1)}, \Omega, \Gamma^{(k, r)})$
        \State $\Gamma_{ij}^{(k, r+1)} \leftarrow \Gamma_{ij}^{(k, r)} + \rho (\Omega_{ij}^{(k, r+1)} - \Theta_i^{(k, r+1)} + \Theta_j^{(k, r+1)})$
        \If{$\|A\Theta^{(k, r+1)}-\Omega^{(k, r+1)}\|_F \leq \epsilon_{\text{pri}}$ \textbf{and} $\|\rho A^\top(\Omega^{(k, r+1)}-\Omega^{(k, r)})\|_F  \leq \epsilon_{\text{dual}}$}
            \State \textbf{break}
        \EndIf
        \State $r \gets r + 1$
    \EndWhile
    \State $\Theta^{(k+1)} \gets \Theta^{(k, r+1)}$ 
    
    \State $\delta_\Theta \gets \|\Theta^{(k+1)}-\Theta^{(k)}\|_F/\|\Theta^{(k)}\|_F$,  $\delta_S \gets \|S^{(k+1)}-S^{(k)}\|_2/\|S^{(k)}\|_2$
    \State $k \gets k + 1$
\Until{$\max(\delta_\Theta, \delta_S) < \epsilon_{\text{rel}}$}
\State \Return ${\Theta}^{(k)},  S^{(k)}$
\end{algorithmic}
\end{algorithm}

\vspace{-10pt}
\subsection{Proximal ADMM algorithm for $\Theta$ update}
We now discuss updating $\Theta$. Without the simplex constraints on the rows of $\Theta$, the update for \( \Theta \) is a generalized convex clustering problem. However, the computational techniques developed for standard convex clustering \citep{chi2015splitting} cannot be straightforwardly applied when the $\theta_i$ are constrained to $\Delta^{K-1}$. Thus, to solve this problem, we propose to use a variation of the alternating direction method of multipliers (ADMM) algorithm. 

The ADMM algorithm can be used to solve convex optimization problems that are the sum of closed proper convex functions. To see how to apply ADMM in our context, we first rewrite the constrained optimization problem as an unconstrained problem, i.e., 
$$ \Theta^{(k+1)} = \argmin_{\Theta \in \mathbb{R}^{n \times K}} \left\{\frac{1}{n}\sum_{i=1}^n \mathcal{F}_i(\theta_i, s_i) + \lambda \sum_{i < j} \gamma_{ij} \|\theta_i - \theta_j\|_2   + \sum_{i=1}^n \delta_{\Delta^{K-1}}(\theta_i) \right\}, $$
where 
$\delta_{\Delta^{K-1}}(\theta_i) =\infty \cdot \mathbf{1}(\theta_i \not\in \Delta^{K-1})$ with the convention $\infty \cdot 0 = 0$.
Evidently, this objective function is the sum of proper closed convex functions.

In order to apply ADMM, we first introduce auxiliary variables $\Omega_{ij} = \theta_i - \theta_j$ for each pair $(i,j)$ that is penalized. The objective function for the $\Theta$ update can then be written as a linearly constrained optimization problem
$$ \argmin_{\Theta, \Omega} \left\{\frac{1}{n}\sum_{i=1}^n \mathcal{F}_i(\theta_i, s_i) + \lambda \sum_{i < j} \gamma_{ij} \|\Omega_{ij}\|_2 + \sum_{i=1}^n \delta_{\Delta^{K-1}}(\theta_i)\right\}$$ $$~~~~~~\text{ subject to }\Omega_{ij} = \theta_i - \theta_j \text{ for all } (i, j) \text{ such that $\gamma_{ij} > 0$}.$$
The augmented Lagrangian for the constrained problem is 
\begin{align*}
\mathcal{L}_\rho(\Theta, \Omega, \Gamma) &= \left\{\frac{1}{n}  \sum_{i=1}^n \mathcal{F}_i(\theta_i, s_i) + \lambda \sum_{i < j} \gamma_{ij} \|\Omega_{ij}\|_2 + \sum_{i=1}^n \delta_{\Delta^{K-1}}(\theta_i)\right.\\
& \qquad\qquad\qquad \left. + \sum_{i < j}  \left[\Gamma_{ij}^\top (\Omega_{ij} - \theta_i + \theta_j) +\frac{\rho}{2} \|\Omega_{ij} - \theta_i + \theta_j\|_2^2 \right] \right\},
\end{align*}
where \( \Gamma_{ij} \) is the Lagrange variable associated with the constraint $\Omega_{ij} = \theta_i - \theta_j$, and \( \rho > 0 \) is a penalty parameter. To be more succinct, define $A$ as the matrix such that \( A_{ij} = [(e_i - e_j)^\top \otimes I_K] \), where $e_i$ is the $i$th basis vector in $\mathbb{R}^n$, so that all $\Omega = A\Theta$ encompasses all constraints $\Omega_{ij} = \theta_i - \theta_j$ for all $(i,j)$ such that $\gamma_{ij} > 0$. Similarly organizing the Lagrange multipliers $\Gamma_{ij}$ into matrix $\Gamma$, we can write the augmented Lagrangian as 
$
\frac{1}{n}  \sum_{i=1}^n\mathcal{F}_i(\theta_i, s_i) + \lambda \sum_{i < j} \gamma_{ij} \|\Omega_{ij}\|_2 + \sum_{i=1}^n \delta_{\Delta^{K-1}}(\theta_i) +  {\rm tr}\left[\Gamma^\top (\Omega - A\Theta)\right] +\frac{\rho}{2} \|\Omega - A\Theta\|_F^2$
where $\|A\|_F^2 = {\rm tr}(A^\top A).$

The standard ADMM algorithm proceeds by minimizing the augmented Lagrangian first with  respect to $\Theta$, then with respect to $\Omega$, then by taking a gradient ascent step with respect to the Lagrange multipliers $\Gamma$. That is, the algorithm has updating equations
\begin{align}
    \Theta^{(r+1)} &= \argmin_{\Theta} \mathcal{L}_\rho(\Theta, \Omega^{(r)}, \Gamma^{(r)})\label{eq:Theta_update}\\
    \Omega^{(r+1)} &= \argmin_{\Omega} \mathcal{L}_\rho(\Theta^{(r+1)}, \Omega, \Gamma^{(r)})\label{eq:Omega_update}\\
    \Gamma^{(r+1)} &= \Gamma^{(r)} + \rho (\Omega^{(r+1)} - A\Theta^{(r+1)}).\notag
\end{align}
In this version of the algorithm, \eqref{eq:Theta_update} can be expressed
$$\argmin_{\Theta} \left\{ \frac{1}{n} \sum_{i=1}^n \mathcal{F}_i(\theta_i, s_i) + \sum_{i=1}^n \delta_{\Delta^{K-1}}(\theta_i) +  {\rm tr}\left[(\Omega^{(r)} - A\Theta)^\top \Gamma^{(r)}\right] +\frac{\rho}{2} \|\Omega^{(r)} - A\Theta\|_F^2  \right\}.$$
This problem is highly nontrivial. First, it involves the simplex constraint. Second, even ignoring the simplex constraint, it is essentially a least squares problem with an additional term $\frac{1}{n} \sum_{i=1}^n \mathcal{F}_i(\theta_i, s_i) =: \mathcal{F}(\Theta,S)$. Third, because the $s_i$ can vary substantially in magnitude, the problem can be severely ill-conditioned.

To handle the simplex constraint, we could use a projected gradient descent sub-algorithm to compute $\Theta^{(r+1)}$. Let $\Theta^-$ be the current iterate of the projected gradient descent subalgorithm. The new iterate of the projected gradient descent subalgorithm is given by
$\Theta^+ = \Pi_{\Delta^{K-1}} (\Theta^- - \alpha \nabla_\Theta \mathcal{L}^0_\rho(\Theta^-, \Omega^{(r)}, \Gamma^{(r)}))$
where $\Pi_{\Delta^{K-1}}:\mathbb{R}^{n \times K} \to \mathbb{R}^{n \times K}$ is the projection operator such that each row of its input is projected onto $\Delta^{K-1}$, $\alpha > 0$ is a step size, and $\mathcal{L}^0_\rho$ is the augmented Lagrangian excluding the term $\sum_{i=1}^n \delta_{\Delta^{K-1}}(\theta_i)$. The gradient of $\mathcal{L}^0_\rho$ with respect to $\Theta$ is 
$ \nabla_\Theta \mathcal{F}(\Theta,S) - A^\top \Gamma + \rho A^\top (\Omega - A\Theta).$
Unfortunately, evaluating this gradient is prohibitively expensive, as it requires the multiplication of $\Theta$ with $A^\top A$ at every iteration, a computation requiring $O(K n^2)$ flops. 

To avoid this expensive computation and to handle ill-conditioning, we modify this approach in two ways. First, we instead approximate \eqref{eq:Theta_update} by minimizing a quadratic majorization of the augmented Lagrangian constructed at $\Theta^{(r)}$. In particular, we define the $(r+1)$-th iterate of our proximal ADMM algorithm as
\begin{align*} 
    \Theta^{(r+1)} &= \argmin_{\Theta} \left\{ \frac{1}{n} \sum_{i=1}^n \mathcal{F}_i(\theta_i, s_i) + \sum_{i=1}^n \delta_{\Delta^{K-1}}(\theta_i)  + {\rm tr}\left[(\Omega^{(r)} - A\Theta)^\top \Gamma^{(r)} \right] \right. \\ 
    &\qquad\qquad \qquad \left. + \frac{\rho}{2} \|\Omega^{(r)} - A\Theta\|_F^2 
  + \frac{\rho}{2}{\rm tr}\left[(\Theta-\Theta^{(r)})^\top (\eta I_n - A^\top A)(\Theta-\Theta^{(r)})\right]  \right\},
\end{align*}
where \( \eta > 0 \) is a constant chosen so that $(\eta I_n - A^\top A)$ is positive semidefinite, e.g., $\eta = \varphi_{\max}(A^\top A)$. It is easy to verify that the objective function above majorizes the augmented Lagrangian at \( \Theta^{(r)} \) because the quadratic term is nonnegative. Thus, $ \Theta^{(r+1)}$ is ensured to produce a decrement of the augmented Lagrangian by the majorize-minimize principle.
The motivation for this particular approximation is that it leads to some helpful cancellations. In particular, the objective function for $\Theta^{(r+1)}$ is
\begin{equation}\label{eq:objective_mid}
\underbrace{\frac{1}{n} \sum_{i=1}^n \mathcal{F}_i(\theta_i, s_i) + \frac{\rho\eta}{2}\|\Theta - \Theta^{(r)} - (\rho\eta)^{-1} A^\top(\rho \Omega^{(r)} - \rho A\Theta^{(r)} + \Gamma^{(r)})\|_F^2}_{=:\mathcal{M}_\rho(\Theta)} + \sum_{i=1}^n \delta_{\Delta^{K-1}}(\theta_i).
\end{equation}
The gradient of $\mathcal{M}_\rho$ is
$
\nabla_\Theta \mathcal{F}(\Theta,S) + \rho\eta (\Theta - \Theta^{(r)}) - A^\top(\rho \Omega^{(r)} - \rho A\Theta^{(r)} + \Gamma^{(r)}).
$
Notice, to apply projected gradient descent to minimize $\mathcal{M}_\rho(\Theta) + \sum_{i=1}^n \delta_{\Delta^{K-1}}(\theta_i)$, we need only compute $A^\top A \Theta^{(r)}$ and $A^\top \Gamma^{(r)}$ once before running our projected gradient descent sub-algorithm: no other multiplications involving $A^\top A$ are needed. This is a substantial computational savings.
Furthermore, the objective function \eqref{eq:objective_mid} is separable across the $\theta_i$. In particular, we can write $\mathcal{M}_\rho(\Theta) = \sum_{i=1}^n \mathcal{M}^i_\rho(\theta_i)$ with
\[
\mathcal{M}^i_\rho(\theta_i) = \frac{1}{n}\sum_{g=1}^G \{ s_i \cdot b_g^\top \theta_i - x_{gi}\log(s_i \cdot b_g^\top \theta_i) \} + \frac{\rho \eta}{2}\left\|\theta_i - \theta_i^{(r)} - \frac{r_i}{\rho\eta}\right\|_2^2,
\]
where $r_i$ is the $i$th row of $A^\top(\rho\Omega^{(r)} - \rho A \Theta^{(r)} + \Gamma^{(r)})$.
Hence, we can compute each $\theta_i^{(r+1)}$ in parallel using a projected gradient descent algorithm with iterates 
$
\theta_i^{+} = \Pi_{\Delta^{K-1}} \{\theta_i^{-} - \alpha \nabla_{\theta_i} \mathcal{M}^i_\rho(\theta_i^{-})\}
$
where $\alpha > 0$ is a step size. This allows for different step sizes for each $\theta_i$ update, making the different scales of the $s_i$ unproblematic. Note that there exist many efficient algorithms for projecting onto the simplex, e.g., that of \citet{condat2016fast}. 

In the proximal ADMM algorithm, the update for $\Omega$ remains as in \eqref{eq:Omega_update}. The update for $\Omega$ can be expressed in the closed form
$
\Omega_{ij}^{(r+1)} = \text{prox}_{\frac{\lambda \gamma_{ij}}{\rho}, \|\cdot\|_2} ( \theta_i^{(r+1)} - \theta_j^{(r+1)} - \rho^{-1} \Gamma_{ij}^{(r)}),
$
where ${\rm prox}_{\phi, \|\cdot\|_2}(a) = \max(1 - \phi/\|a\|_2, 0)a$ for vector $a$ and nonnegative scalar $\phi$.  Applying results from \citet{deng2016global}, this approach has roughly the same convergence guarantees as the vanilla ADMM algorithm.

We use the convergence criterion for the ADMM algorithm from \citet{boyd2011distributed}, where the proximal ADMM is said to have converged if
$
\|k^{(r)}\| \leq \epsilon^{\rm{pri}}, \|s^{(r)}\| \leq \epsilon^{\rm{dual}}$,
where
$
k^{(r)} = A\Theta^{(r)}-\Omega^{(r)}$ and $s^{(r)} = \rho A(\Omega^{(r-1)} - \Omega^{(r)}).
$
In theory, both primal and dual residuals will converge to zero for any $\rho > 0$. However, in practice, varying the penalty parameter as suggested by \citet{boyd2011distributed} can help speed up convergence. We use the approach of \citet{zhu2017augmented}, which accounts for the magnitude of the tolerances, by updating the penalty parameter according to
$
\rho = 
\tau \rho \mathbf{1}(\|k^{(r)}\|/ \epsilon^{\rm{pri}} \geq \mu \|s^{(r)}\|/\epsilon^{\rm{dual}}) + 
\tau^{-1}\rho \mathbf{1}(\|s^{(r)}\|/\epsilon^{\rm{dual}} \geq \mu \|k^{(r)}\|/\epsilon^{\rm{pri}}),$
where by default, we set $\tau = 2$ and $\mu = 10$.
\section{Implementation details}
\label{s:implementation}
\subsection{Spatially-informed weighting scheme}\label{sec:weight_construction}
To incorporate spatial dependence in the clustering penalty, we construct a weight matrix using similarities between the cell-type compositions of neighboring spots. Specifically, we first construct $\tilde{\theta}_i \in \Delta^{K-1}$, the local average of the $k^*$-nearest adjacent neighbors' pilot estimates of the cell-type proportion in spot $i$ obtained from spot-by-spot deconvolution. More precisely, let $\widebar{\theta}_i$ be an estimate of the $i$th spot's cell-type proportion, defined in \eqref{eq:spot-by-spot}. Then, for all spots $j$ adjacent to spot $i$, we compute and sort $\|\widebar{\theta}_i - \widebar{\theta}_j\|_2$ according to their magnitude. We then define $\mathcal{K}_i$ as the set of spots adjacent to spot $i$ whose $\|\widebar{\theta}_i - \widebar{\theta}_j\|_2$ was among the $k^*$ smallest. We formally define 
$\tilde{\theta}_i = \frac{1}{k^*}\sum_{j \in \mathcal{K}_i} \widebar{\theta}_j.$ With the $\tilde{\theta}_i$, we define the weights between two adjacent spots \( i \) and \( j \) as
$$
\gamma_{ij} = \left\{\begin{array}{ll}
    \exp\left( -\frac{\| \tilde{\theta}_i - \tilde{\theta}_j \|_2^2}{2\sigma_i \sigma_j} \right) & :\text{spot $i$ and $j$ are adjacent}\\
    0 & :\text{otherwise}
\end{array}\right..
$$
The $\sigma_i$ and $\sigma_j$ are adaptive bandwidth parameters inspired by \citet{chi2025and}. Specifically, $\sigma_i$ is the median of the $k^{**}$ smallest $\|\tilde{\theta}_i - \tilde{\theta}_j\|_2$ over all spots $j$ adjacent to spot $i$. By default, we take $k^* = k^{**} = 5$. In Section 3 of the Supplementary Materials, we show that our method is quite robust to these choices. 

Intuitively, the weight between spots $i$ and $j$ is a decreasing function of 
  the distance between their neighborhood-averaged realized cell-type
  proportions, $\tilde\theta_i$ and $\tilde\theta_j$. The rationale is that if  
  $\tilde\theta_i$ and $\tilde\theta_j$ are close, it is reasonable to assume
  the expected cell-type proportions $\theta_i^*$ and $\theta_j^*$ are also close,
  so the penalty term $\|\theta_i - \theta_j\|_2$ should be weighted more heavily.

\begin{figure}[t]
    \centering
    \begin{subfigure}{0.32\textwidth}
        \centering
        \includegraphics[width=\textwidth]{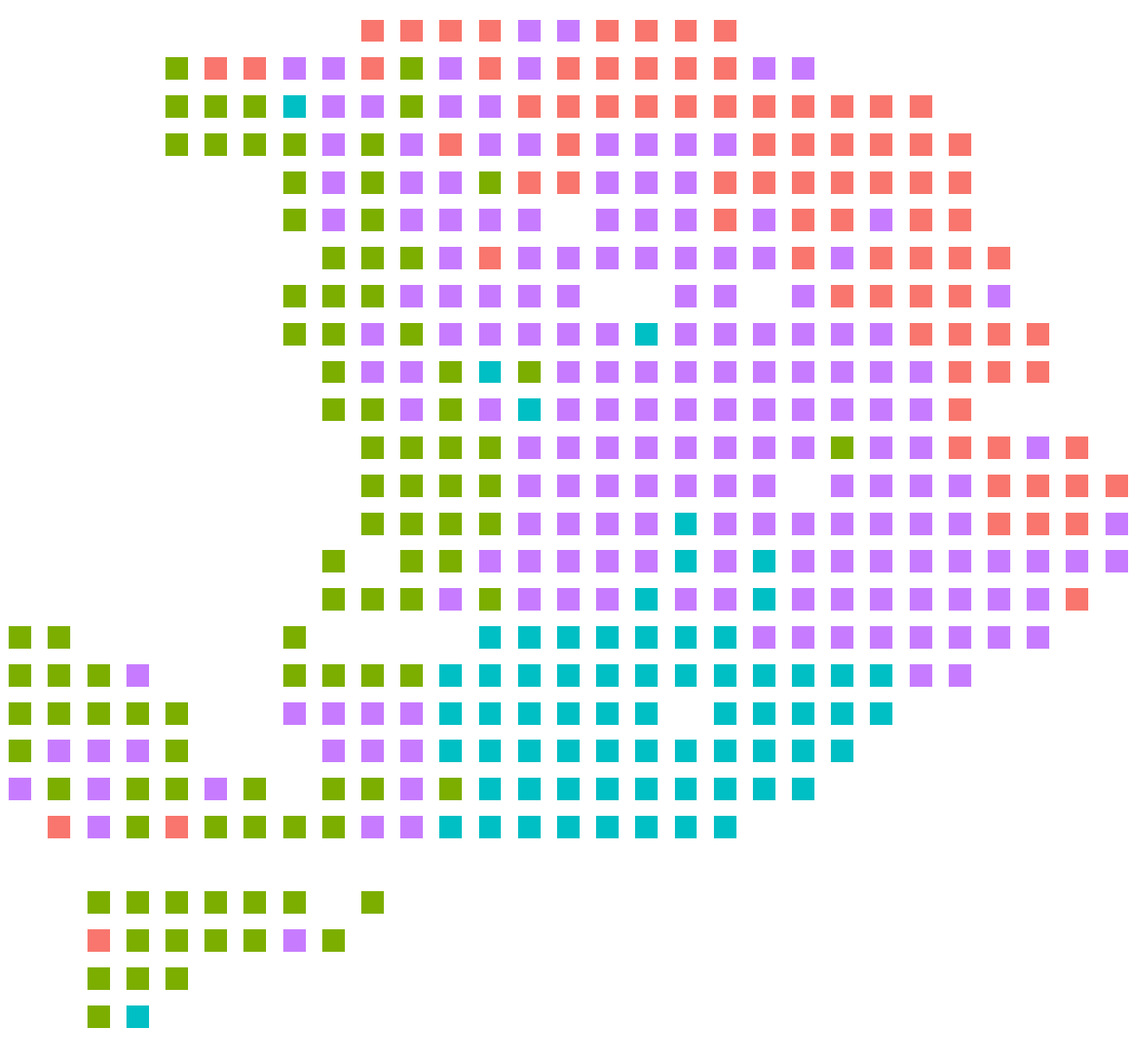}
        \caption{Histological annotation}
        \label{fig:sub1}
    \end{subfigure}\hfill
    \begin{subfigure}{0.32\textwidth}
        \centering
        \includegraphics[width=\textwidth]{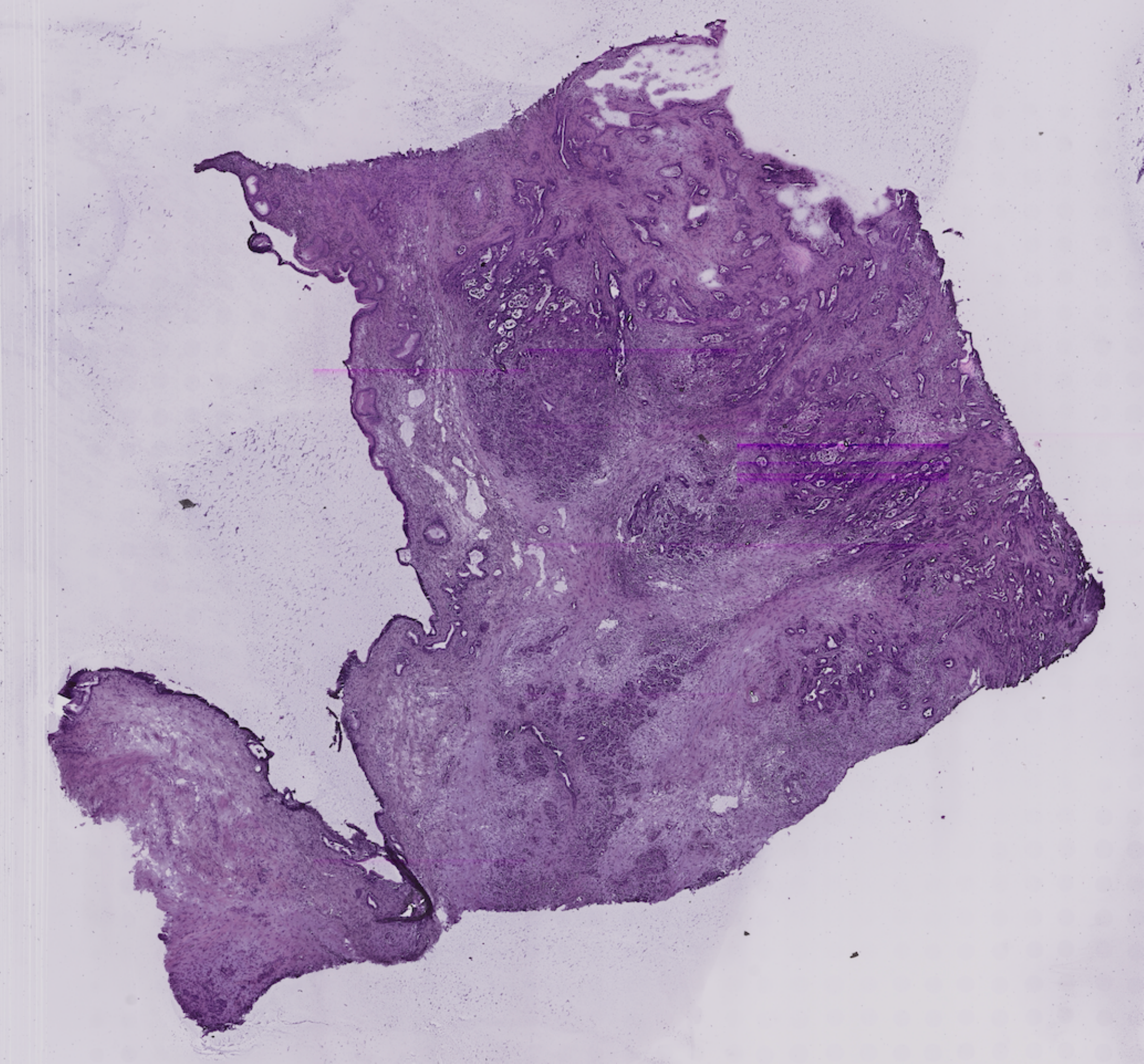}
        \caption{Histological image}
        \label{fig:sub2}
    \end{subfigure}\hfill
    \begin{subfigure}{0.32\textwidth}
        \centering
        \includegraphics[width=\textwidth]{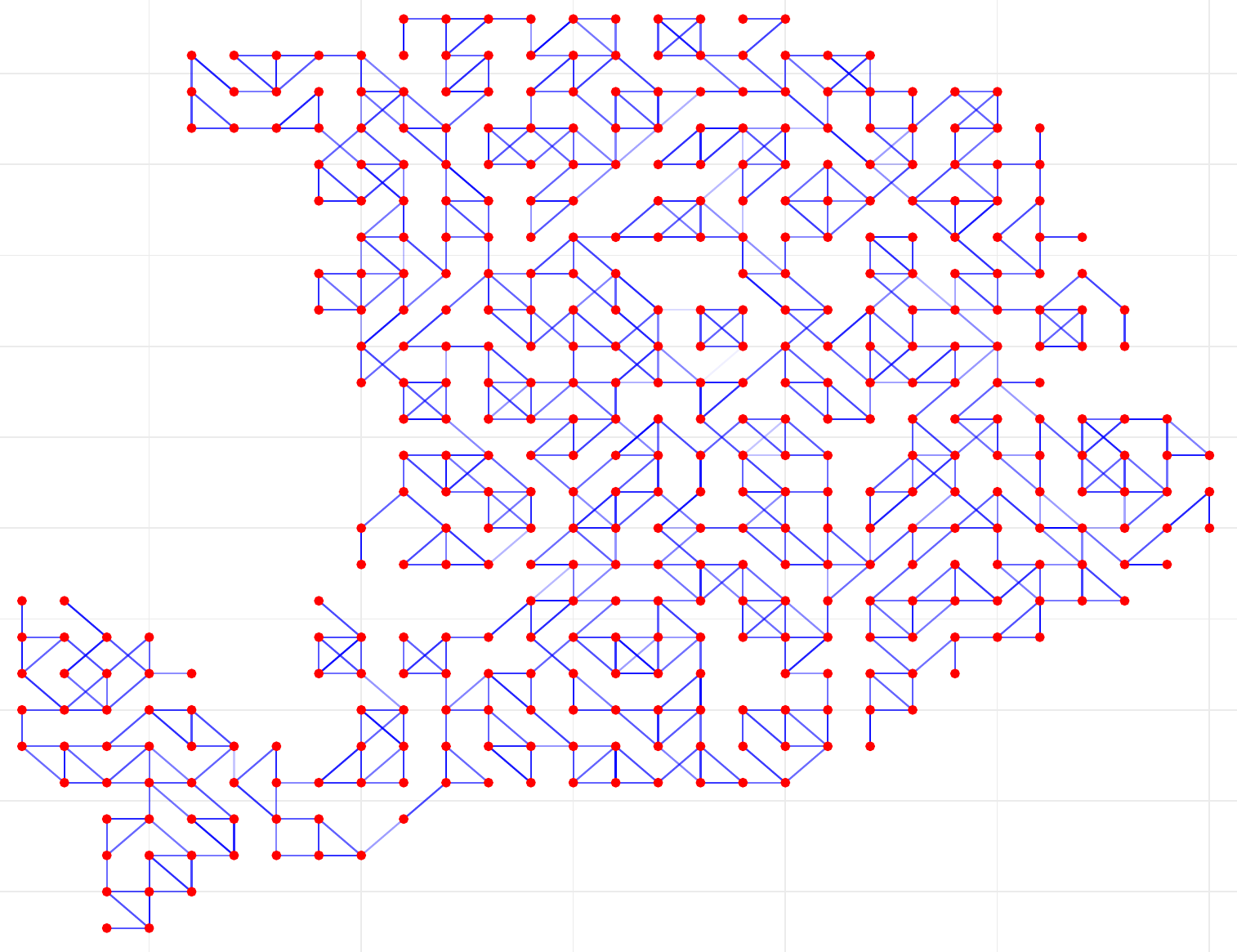} 
        \caption{Spatial weights} 
        \label{fig:sub3}
    \end{subfigure}
    \caption{PDAC histological image and annotations. Panel (a) shows the histological annotations with colors representing different tissue types: red as cancer, green as ductal epithelium, purple as stroma, and blue as normal pancreatic tissue. Panel (b) is the corresponding H\&E-stained histological image of the PDAC tissue section. Panel (c) shows the spatial weights $\gamma_{ij}$ between adjacent spots, where edge thickness represents the magnitude of the weights (i.e. thicker edges indicate stronger connections).}
    \label{fig:PDAC_HE}
\end{figure}

To sparsify the weight matrix while preserving connectivity, we use an iterative approach where we first prune the weights by removing some proportion of connections for each individual node (e.g., we use 10\% in simulations and the PDAC analyses). Then to ensure connectivity, we add a minimum spanning tree (MST) on the pruned weights. The MST is computed on the graph defined by spatial adjacency and Gaussian kernels, ensuring that each spot remains connected to at least one neighbor. This sparsification reduces the computational burden while retaining important spatial relationships. Finally, we add a lower bound on the nonzero weights to avoid numerical instability in optimization. For an example of a weight matrix constructed in this way, see Figure \ref{fig:PDAC_HE}(c).

\subsection{Initialization for solution path computation}
\label{sec:initialization}
In this section, we first establish the smallest tuning parameter value $\lambda^*$ for which all spatial spots collapse into a single global domain. This can be used to initialize the solution path computation. Let $D \in \mathbb{R}^{E \times n}$ be the matrix such that $D\Theta$ is the matrix of weighted pairwise differences $\gamma_{ij}(\theta_i - \theta_j)$ for all pairs  $(i,j)$ such that $\gamma_{ij} > 0$. In other words, $D$ is the weighted incidence matrix of the graph defined by the nonzero weights $\gamma_{ij}$. 
\begin{prop}
\label{prop:lambda_trivial}
Define $\widebar\theta$ as the solution to the single-domain problem, i.e., \\
$(\widebar{\theta}, \widebar{s}) = \argmin_{\theta \in \Delta^{K-1}, s_i \geq 0} \sum_{i=1}^n \mathcal{F}_i(\theta, s_i)$. Let $\widebar{\Theta} = \mathbf{1}_n \widebar{\theta}^\top$ and let 
$G = \nabla_\Theta \frac{1}{n} \sum_{i=1}^n \mathcal{F}_i(\widebar{\theta},\widebar{s}_i) \in \mathbb{R}^{n \times K}$ denote the
gradient of the loss with respect to $\Theta$ evaluated at $\widebar{\Theta}$.
If the weighted incidence matrix $D$ encodes
a connected graph, then $\widebar\Theta$ is the solution to \eqref{eq:general_estimator0} with $S$ fixed at $\widebar{S}$ for all $\lambda \geq \lambda_\text{\rm trivial}$, where
\begin{equation}\label{eq:lambda_trivial}
    \lambda_\text{\rm trivial} = \min_{{V, \eta, \nu}}
    \|V\|_{2,\infty}
    \ \text{subject to} \
    D^\top V = -G - \eta \mathbf{1}_K^\top + \nu,
    \ \nu \geq 0,
    \ \nu_{ik} = 0 \text{ on } \mathrm{supp}(\widebar{\theta}),
\end{equation}
with $\|V\|_{2,\infty} := \max_{\ell \in [E]} \|V_{\ell\cdot}\|_2$. Moreover, \eqref{eq:lambda_trivial} is a second-order cone program \textnormal{(SOCP)}
and attains its minimum.
\end{prop}
A proof of Proposition \ref{eq:lambda_trivial} can be found in the Supplementary Materials. 

To evaluate $\lambda_{\mathrm{trivial}}$ in practice, we must first compute the trivial 
single-domain solution $\widebar{\Theta} = \mathbf{1}_n\widebar{\theta}^\top$ and its associated 
size factors $\widebar{S} = \mathrm{diag}(\widebar{s}_1, \dots, \widebar{s}_n)$. Here, we focus on the Poisson case for concreteness. When all rows of 
$\Theta$ are constrained to be equal to a common vector $\widebar{\theta} \in \Delta^{K-1}$, 
the DUET objective reduces to 
\begin{equation}\label{eq:trivial_obj}
  \min_{\widebar{\theta} \in \Delta^{K-1},\, s_i \geq 0}
  \frac{1}{n}
  \sum_{i=1}^{n}\sum_{g=1}^{G}
  \bigl\{
    s_i\,b_g^\top\widebar{\theta}
    - x_{gi}\log(s_i\,b_g^\top\widebar{\theta})
  \bigr\},
\end{equation}
since the pairwise penalty vanishes identically. This is a rank-one Poisson 
factorization problem, as the fitted mean matrix $B\widebar{\theta}\,s^\top \in \mathbb{R}^{G \times n}$ 
is rank one. We solve \eqref{eq:trivial_obj} via a two-step alternating algorithm. With 
$\widebar{\theta}$ fixed, the objective separates over $i$ and the exact minimizer of each 
$s_i$ is obtained by setting the partial derivative to zero, i.e., 
  $\widebar{s}_i \leftarrow
 \left(\sum_{g=1}^G b_g^\top\widebar{\theta}\right)^{-1} \sum_{g=1}^G x_{gi} ,$
 for $i \in [n].$
With $\widebar{S}$ fixed, we update $\widebar{\theta}$ via the Lee--Seung multiplicative update 
\citep{lee1999learning}, which is essentially a majorize-minimize algorithm that monotonically decreases the 
objective in $\widebar{\theta}$ \citep{hunter2004tutorial}. Letting 
$\widebar{S} = \sum_{i=1}^n \widebar{s}_i$ and 
$x_{g\cdot} = \sum_{i=1}^n x_{gi}$, the update for each component $k \in [K]$ is
\begin{equation}\label{eq:theta_trivial}
  \widebar{\theta}_k \leftarrow
  \widebar{\theta}_k \cdot
  \frac{\displaystyle\sum_{g=1}^G
        b_{gk}\, x_{g\cdot} \big/ (b_g^\top\widebar{\theta})}
       {\widebar{S}\,\displaystyle\sum_{g=1}^G b_{gk}}.\notag
\end{equation}
The two steps are iterated until the objective function value converges. Upon 
convergence, we normalize $Z = \sum_k \widebar{\theta}_k$ and set 
$\widebar{\theta} \leftarrow \widebar{\theta}/Z$, $\widebar{s}_i \leftarrow \widebar{s}_i \cdot Z$ 
for all $i$, which imposes $\widebar{\theta} \in \Delta^{K-1}$ without altering any fitted 
mean $\widebar{s}_i\,b_g^\top\widebar{\theta}$. 


\subsection{Cluster assignment}
\label{sec:postprocessing}
After the proximal ADMM algorithm converges, the cluster structure is encoded in the auxiliary variables $\{\Omega_{ij}\}$. When $\Omega_{ij} = \mathbf{0}$, the constraint $\theta_i = \theta_j$ is active, indicating that spots $i$ and $j$ belong to the same spatial domain.  We recover the partition by applying a union-find algorithm for each penalized pair $(i,j)$ with $\Omega_{ij}=\mathbf{0}$, the spots $i$ and $j$ are unioned.  The resulting connected components define the estimated domains, and all spots within a component are assigned the component centroid $n_c^{-1}\sum_{i\in\mathcal{C}_c}\hat\theta_i$ as their common composition estimate.

Due to finite numerical precision, the ADMM solution may produce $\hat\theta_i$ values that are nearly but not exactly equal across spots.  Following the convention of \citet{chi2015splitting}, we apply a hard-thresholding step after union-find: spots $i$ and $j$ are additionally merged if
$
  \|\hat\theta_i - \hat\theta_j\|_2
    < \widehat\sigma_d/4
$
where $\widehat\sigma_d$ is the empirical standard deviation of all pairwise distances $\{\|\hat\theta_i-\hat\theta_j\|_2 : i < j\}$.
This data-adaptive threshold calibrates fusion to the scale of variation in the estimated compositions.  Union-find and centroid averaging are applied again to the merged graph, yielding the final hard-clustered estimate $\widehat\Theta^{\lambda}$.

\subsection{Tuning parameter selection with BIC}
A computationally efficient way to select the tuning parameter is to use the Bayesian Information Criterion (BIC).  Let $\ell_j$ be the log-likelihood for the $j$th tuning parameter, and let $c_j$ be the number of clusters returned.  Then we define
\begin{equation}
\label{eq:bic}
\mathrm{BIC}_j = -2 \ell_j + \{(K - 1) c_j + n\} \log(n),
\end{equation}
where $n$ is the total number of spots, and $K$ is the total number of cell types. We approximate the degrees of freedom using $(K - 1) c_j + n$: $(K - 1) c_j$ is the number of cell-type composition parameters estimated by our method, and $n$ is the number of size factor parameters. 

To efficiently compute \eqref{eq:general_estimator} over a candidate grid $\lambda_1 \geq \lambda_2 \geq \dots \geq\lambda_L$, we fit the model sequentially and use warm starts: the algorithm for solving \eqref{eq:general_estimator} at $\lambda_{k-1}$ is initialized at the solution to \eqref{eq:general_estimator} at $\lambda_k$ for each $k \in \{2, \dots, L\}$. In our implementation, $\lambda_1 = \lambda_{\mathrm{trivial}}$.

\section{Simulations}\label{sec:Simulations}
\label{s:simulations}
\subsection{Data generating models}
\begin{figure}[t!]
\centering
\includegraphics[width=0.9\textwidth]{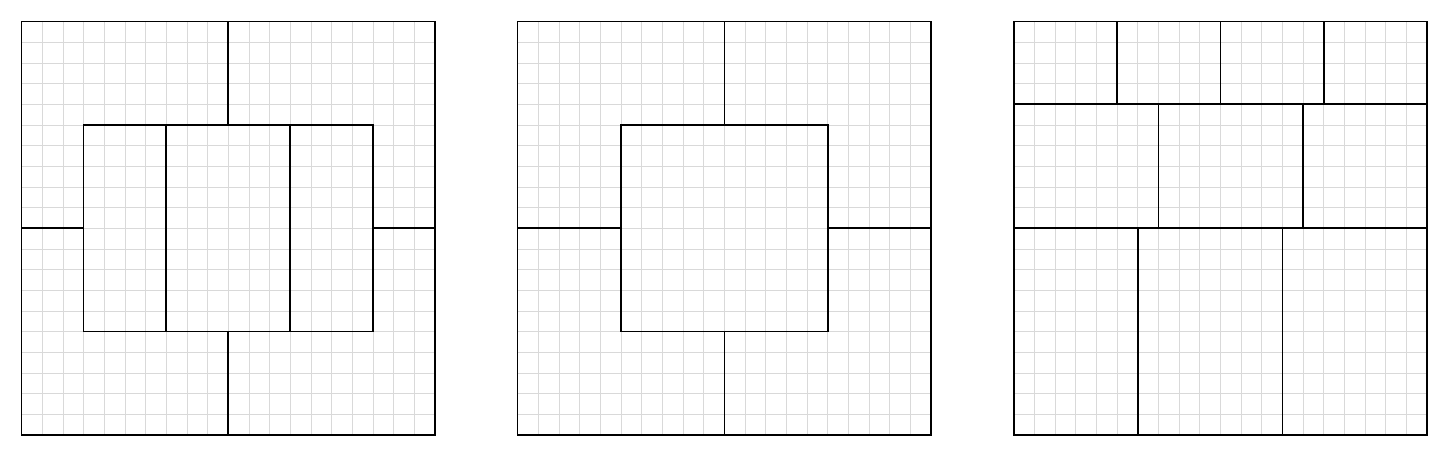}
 \caption{Cluster structures for the simulation studies with (left) 5 clusters, (center) 7 clusters, and (right) 10 clusters. Each color represents a different spatial cluster.}\label{fig:sim_clusters}
    \label{fig:three_figures}
\end{figure}

To assess the performance of our method, we conducted comprehensive simulation studies under different numbers of clusters and smoothness settings. For each model and setting, we performed 100 independent replications. In each of the replications, we generated a $20 \times 20$ spatial grid comprising $n = 400$ spatial spots (e.g., see Figure \ref{fig:sim_clusters}).

For each replication, we constructed an expected cell-type composition matrix $\Theta^* = (\theta^*_1, \dots, \theta^*_n)^\top$, where $\theta^*_i \in \Delta^{K-1}$ with $K = 5$ distinct cell types. The $i$th row of $\Theta^*$, $\theta^*_i$, corresponds to the $i$th spot, and is set equal to one of the rows of a reference matrix $\Theta^{\ddagger}_{(1)}, \Theta^{\ddagger}_{(2)}, \text{ or } \Theta^{\ddagger}_{(3)}$, depending on the scenario considered. Specifically, all spots within a given spatial cluster share the same characteristic expected cell-type composition vector, $\theta^\ddagger$,  corresponding to a particular row of $\Theta^\ddagger$. To be precise, we set 
$
\theta^*_{i\cdot} = [\Theta^\ddagger_{(m)}]_{\mathcal{C}(i) \cdot}$ {for all spots } $i \in [n]$
where $\mathcal{C}:[n] \to \{1, \dots, C\}$ is the function mapping a spot to the index of its spatial domain, $C$ is the number of distinct spatial domains, and $[\Theta^\ddagger_{(m)}]_{j \cdot}$ is the $j$th row of $\Theta^\ddagger_{(m)}$. The matrices $\Theta^{\ddagger}_{(1)}, \Theta^{\ddagger}_{(2)}, \text{ and } \Theta^{\ddagger}_{(3)}$ represent three different levels of smoothness where $\Theta^\ddagger_{(1)}$ reflects the setting with the most well-separated cell types with cluster-specific dominance, while $\Theta^\ddagger_{(2)}$ and $\Theta^\ddagger_{(3)}$ introduce more overlapping cell-type compositions across clusters. For the $C=5$ case, the expected cell-type composition matrices were defined as follows:
\[
\Theta^\ddagger_{(1)} = {\small\begin{pmatrix}
0.05 ~& 0.05 ~& 0.05 ~& 0.00 ~& 0.85 \\
0.05 ~& 0.05 ~& 0.00 ~& 0.85 ~& 0.05 \\
0.05 ~& 0.00 ~& 0.85 ~& 0.05 ~& 0.05 \\
0.00 ~& 0.85 ~& 0.05 ~& 0.05 ~& 0.05 \\
0.85 ~& 0.05 ~& 0.05 ~& 0.05 ~& 0.00 \\
\end{pmatrix}}, \quad\quad
\Theta^\ddagger_{(2)} = {\small\begin{pmatrix}
0.8 ~& 0.1 ~& 0.1 ~& 0.0 ~& 0.0 \\
0.1 ~& 0.6 ~& 0.3 ~& 0.0 ~& 0.0 \\
0.2 ~& 0.2 ~& 0.3 ~& 0.2 ~& 0.1 \\
0.0 ~& 0.0 ~& 0.1 ~& 0.1 ~& 0.8 \\
0.2 ~& 0.4 ~& 0.0 ~& 0.1 ~& 0.3 \\
\end{pmatrix}}, ~
\]
\[
\Theta^\ddagger_{(3)} = {\small\begin{pmatrix}
0.70 ~& 0.15  ~& 0.10  ~& 0.03 ~& 0.02 \\
0.20 ~& 0.40  ~& 0.10  ~& 0.10 ~& 0.20 \\
0.02 ~& 0.08  ~& 0.85 ~&0.03 ~& 0.02 \\
0.10 ~& 0.25  ~& 0.25 ~& 0.30 ~& 0.10 \\
0.15 ~& 0.05  ~& 0.30 ~& 0.20 ~& 0.30 \\
\end{pmatrix}}.
\]

In practice, the realized composition within a spot will deviate from the underlying expected composition due to sampling variability. To reflect this, we generated the realized composition as a multinomial with $s^*_i$ trials and probability vector $\theta^*_{i}$. In particular, for each spot $i$, we drew from a multinomial distribution where $u_{i} \sim \text{Multinomial}(s^*_i, \theta^*_{i})$, where the spot-specific size factor $s^*_i$ was drawn from the discrete uniform distribution on $[50]$. The realized compositions were then calculated by row-normalizing $v_{i} = u_{i} / s^*_i$.

Finally, the observed spatial transcriptomics count matrix was generated so that $X_{gi}\mid V_i = v_i \sim \text{Poisson}(s^*_i \cdot b_g^\top v_{i})$.  The single-cell reference matrix $B \in \mathbb{R}_+^{66 \times 5}$ was taken from a real mouse olfactory bulb (MOB) dataset described in the Supplementary Materials Section 4. 

Spots were assigned to $C = 5$, $C=7$, or $C=10$ spatial clusters. For a given $C$, the cluster structure was defined by a fixed geometric partition of the grid which can be found in Figure \ref{fig:three_figures}. For the $C=7$ and $C=10$ cases, the matrices $\Theta^\ddagger_{(1)}, \Theta^\ddagger_{(2)}, \Theta^\ddagger_{(3)}$ were defined similarly to those above, with each row representing a distinct cluster's expected cell-type composition. The specific values for these matrices can be found in the Supplementary Materials Section 2. 

We constructed spatial weights as described in Section \ref{sec:weight_construction}, with $k^* = 5$, $k^{**} = 5$ across all settings to match the real data application. We demonstrate the robustness to the number of neighbors used in our weight construction in the Supplementary Materials Section 3.

\subsection{Methods for comparison}
We compared DUET with several existing methods for deconvolution and spatial clustering. For DUET, we used two variations: \underline{DUET}, where we update both $S$ and $\Theta$ using the proposed two-block coordinate descent algorithm; and \underline{App-DUET}, for approximate DUET, where we first estimate $S^*$ using the spot-by-spot approach, then only update $\Theta$ using the proximal ADMM algorithm. The latter approach is essentially an approximation to \eqref{eq:general_estimator0}: it is much faster to compute and can work reasonably well when the spot-by-spot estimate of $s_i^*$ is accurate. 

Other competitors included for comparison are \underline{O-IRIS} \citep{ma2024accurate}, the aforementioned method \eqref{eq:IRIS_criterion} that performs both deconvolution and clustering, with $R$ set to the ``oracle'' number of clusters; \underline{Seurat} \citep{hao2024dictionary}, a widely used software for spatial transcriptomics data analysis that can perform clustering of spots; \underline{O-BayesSpace} for ``oracle" BayesSpace \citep{zhao2020bayesspace}, a Bayesian method for spatial clustering and deconvolution which is given oracle knowledge of the correct number of clusters; \underline{SPOTlight} \citep{elosua2021spotlight}, a method that uses reference scRNA-seq data to deconvolve spatial transcriptomics data; and \underline{CARD} \citep{ma2022spatially}, a method which employs a reference-based approach for spatial deconvolution while leveraging spatial information. We used the oracle versions of the various methods because, to the best of our knowledge, they do not have default approaches for selecting the number of clusters.  Finally, we also compare to spotwise deconvolution, \underline{Spotwise}, which does not do clustering, but does estimate the $\theta_i^*$.
For {Seurat} and {BayesSpace}, we first performed clustering using the gene expression data to identify the spatial domains, then performed the deconvolution on each of the identified clusters. For the two deconvolution methods (SPOTlight and CARD), we first performed deconvolution to estimate the cell-type proportions for each spot, then performed clustering on the estimated proportions by assigning cluster assignments based on the dominant cell type.  Neither {SPOTlight} nor {CARD} performs clustering directly: clustering is done post-hoc. 

\subsection{Results}

\begin{figure}[!t]
    \centering
    \includegraphics[width=1\textwidth]{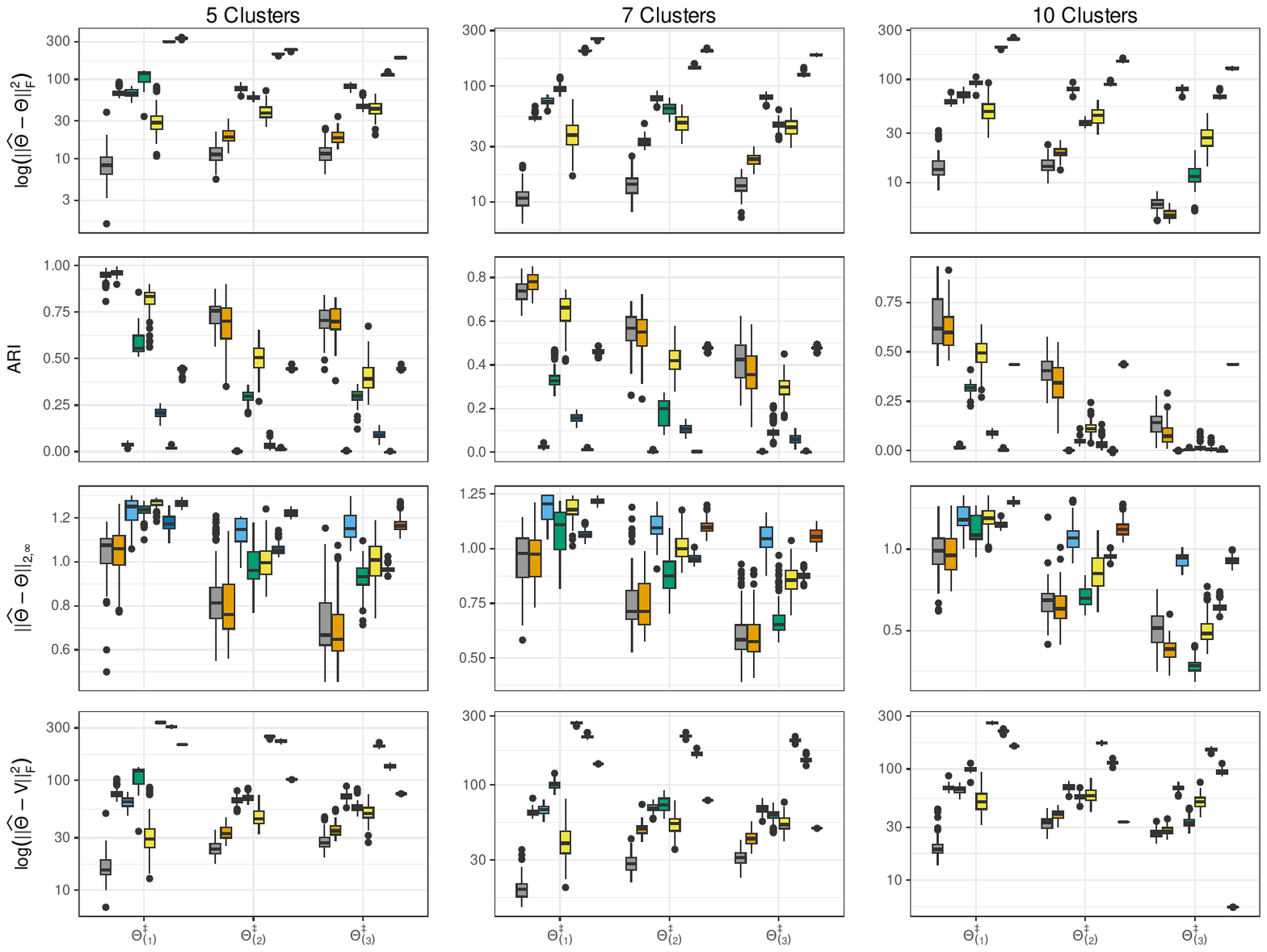}\\
    \includegraphics[width=0.45\textwidth]{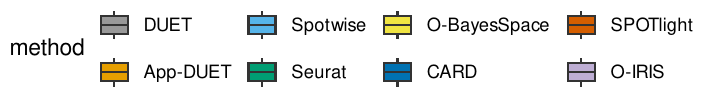}
    \caption{Simulation results comparing our method with existing approaches across 9 different settings defined by the combination of the degree of smoothness (i.e., \(\Theta^\ddagger_{(1)}\), \(\Theta^\ddagger_{(2)}\), and \(\Theta^\ddagger_{(3)}\)) and the number of clusters (i.e., $C \in \{5,7,10\}$). The first row shows the log estimation error ($\log \|\hat{\Theta} - \Theta^*\|^2_F$) for each method. The second row shows the adjusted Rand index (ARI) values, the third row shows maximal row-wise error ($\| \hat{\Theta} - \Theta^* \|_{2,\infty}$), and the fourth row shows the log realized error (\(\log \|\hat{\Theta} - V\|^2_F\)) for each method.}
    \label{fig:sim_results}
\end{figure}

We evaluated our method against competitors across 100 independent replicates for nine experimental settings, varying the degree of spatial smoothness (\(\Theta^\ddagger_{(1)}\), \(\Theta^\ddagger_{(2)}\), \(\Theta^\ddagger_{(3)}\)) and the number of clusters (\(C \in \{5,7,10\}\)). Figure~\ref{fig:sim_results} displays the logarithm of total estimation error (\(\log \| \hat{\Theta} - \Theta^* \|^2_F \)), clustering adjusted Rand index (ARI), maximal row-wise error (\(  \| \hat{\Theta} - \Theta^* \|_{2,\infty} \)), and logarithm of realized composition error (\(\log \| \hat{\Theta} - V \|^2_F \)). Note that CARD, SPOTlight, and IRIS target the unobservable realized proportions ($v_i$), whereas DUET, BayesSpace, and Seurat target the expected proportions ($\theta_i^*$). Consequently, for IRIS, we only report ARI and realized error; for the remaining methods, we report all metrics, treating their cell-type proportion estimates as $\hat\Theta$.

Across all cluster settings ($C \in \{5,7,10\}$), DUET consistently achieved the lowest total estimation error (MSE), clearly outperforming all competitors---including oracle BayesSpace, which has the distinct advantage of knowing the true number of clusters. This performance advantage was evident across all spatial smoothness settings ($\Theta^\ddagger_{(1)}$, $\Theta^\ddagger_{(2)}$, $\Theta^\ddagger_{(3)}$), with DUET maintaining a substantial lead as the number of clusters grew. Furthermore, DUET exhibited excellent robustness to localized composition extremes, consistently yielding among the lowest maximal row-wise errors across all scenarios.

In terms of domain detection, increasing spatial smoothness predictably reduced ARI scores across most methods as cell-type compositions became more homogeneous. Nevertheless, DUET maintained superior clustering accuracy. It noticeably outperformed oracle BayesSpace under lower smoothness ($\Theta^\ddagger_{(1)}$) and remained the top-performing method under higher smoothness ($\Theta^\ddagger_{(3)}$), exceeding the accuracy of oracle IRIS. Furthermore, while DUET explicitly targets the expected compositions ($\theta_i^*$), it achieved the lowest realized errors of all methods evaluated, confirming that accurate expected estimates effectively and reliably recover the realized proportions.

Finally, our computationally efficient variant, approximate DUET, yielded performance nearly identical to the full DUET model across all metrics. This demonstrates that approximate DUET is a highly practical alternative when reasonable size factor estimates are available, though jointly estimating size factors and cell-type proportions (as in full DUET) offers slight, incremental gains in accuracy.

\section{Spatially-resolved transcriptomic data analyses}\label{sec:PDAC}
\subsection{PDAC data description and preparation}
We applied our method to a human pancreatic ductal adenocarcinoma (PDAC) dataset to assess its performance in a real dataset. We utilized the paired SRT and scRNA-seq data from the PDAC-A sample published by \citet{moncada2020integrating}. The spatial data provides gene expression measurements across spatially barcoded oligo-dT microarrays on a 2D tissue section.  The reference scRNA-seq data was generated from the same individual (PDAC-A) using the inDrop platform, providing a matched, high-resolution profile of the major cell types present in the tumor microenvironment, including ductal cells, immune populations, and cancer-associated fibroblasts.

To construct a cell-type reference matrix $B$, we preprocessed the scRNA-seq data as follows. The raw gene-by-cell UMI count matrix was imported, with cell-type annotations extracted from accompanying metadata. We created a {Seurat} object, applying basic quality control by retaining cells expressing at least five genes. We then identified marker genes by performing differential expression analysis across cell types using the Wilcoxon rank-sum test, retaining genes with an adjusted $p$-value $<$ 0.05, $\log_2$ fold change $>$ 0.25, and expression in more than 10\% of cells within a cluster. For each cell type, we computed the mean expression across its constituent cells using raw counts, restricted to the selected marker genes.  Genes with zero expression across all cell types were excluded. For the scRNA-seq data, we consolidated cell-type annotations by merging related subpopulations (e.g., macrophage and ductal subtypes). 
The spatial data were filtered for quality control, filtering out genes with low counts across spatial spots ($<10$) and also spots with low total expression ($ < 100$). After intersecting common genes, the final dataset comprised $G = 2584$ genes across $K = 15$ cell types and $n = 427$ spatial spots, each with associated spatial coordinates and histological annotations.

\begin{figure}[t!]
    \centering
    \begin{subfigure}{0.32\textwidth}
        \centering
        \includegraphics[width=1\textwidth]{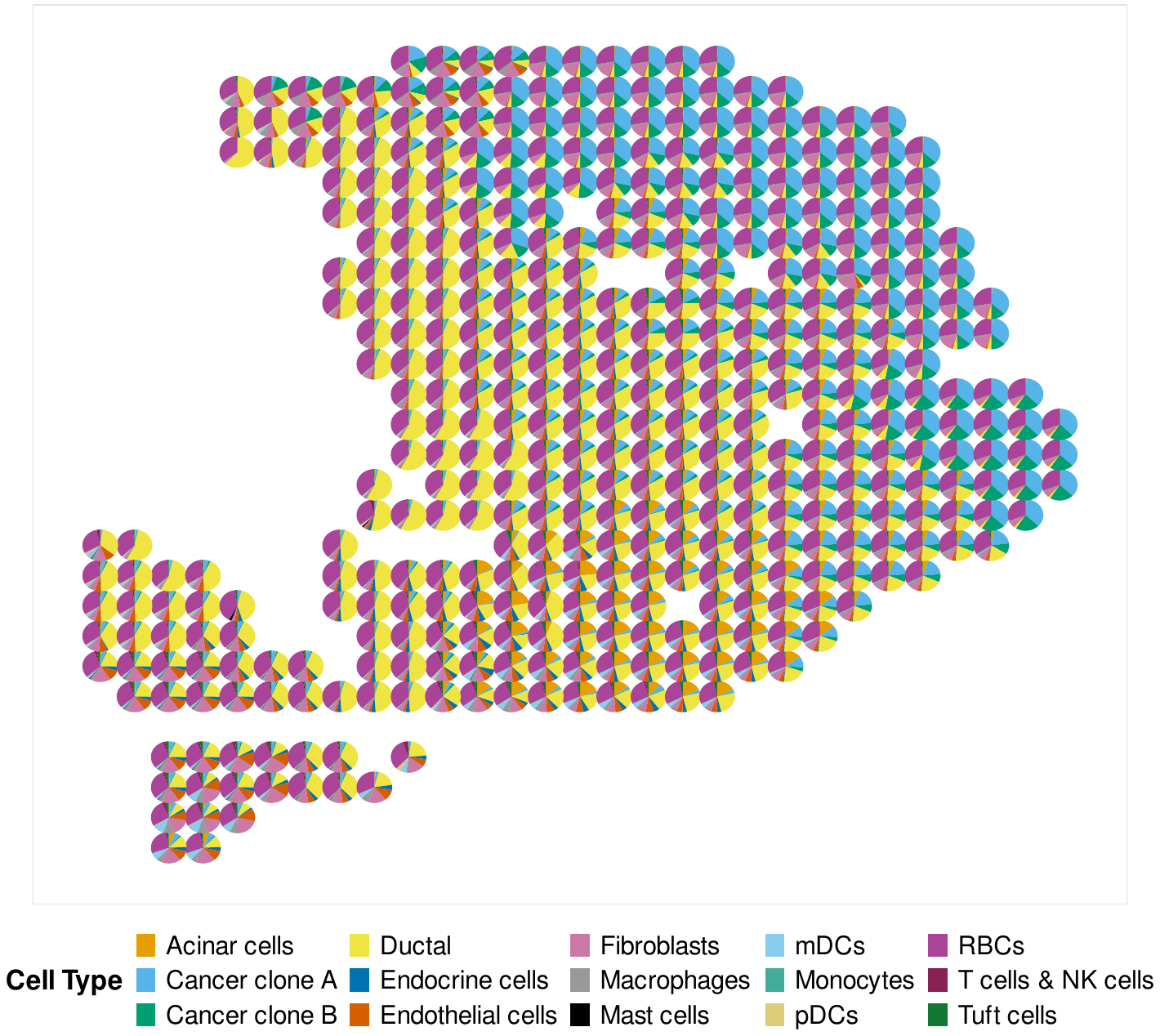} 
        \caption{$\widehat\Theta$: DUET} 
        \label{fig:sub5}
    \end{subfigure}\hfill
    \begin{subfigure}{0.32\textwidth}
        \centering
        \includegraphics[width=1\textwidth]{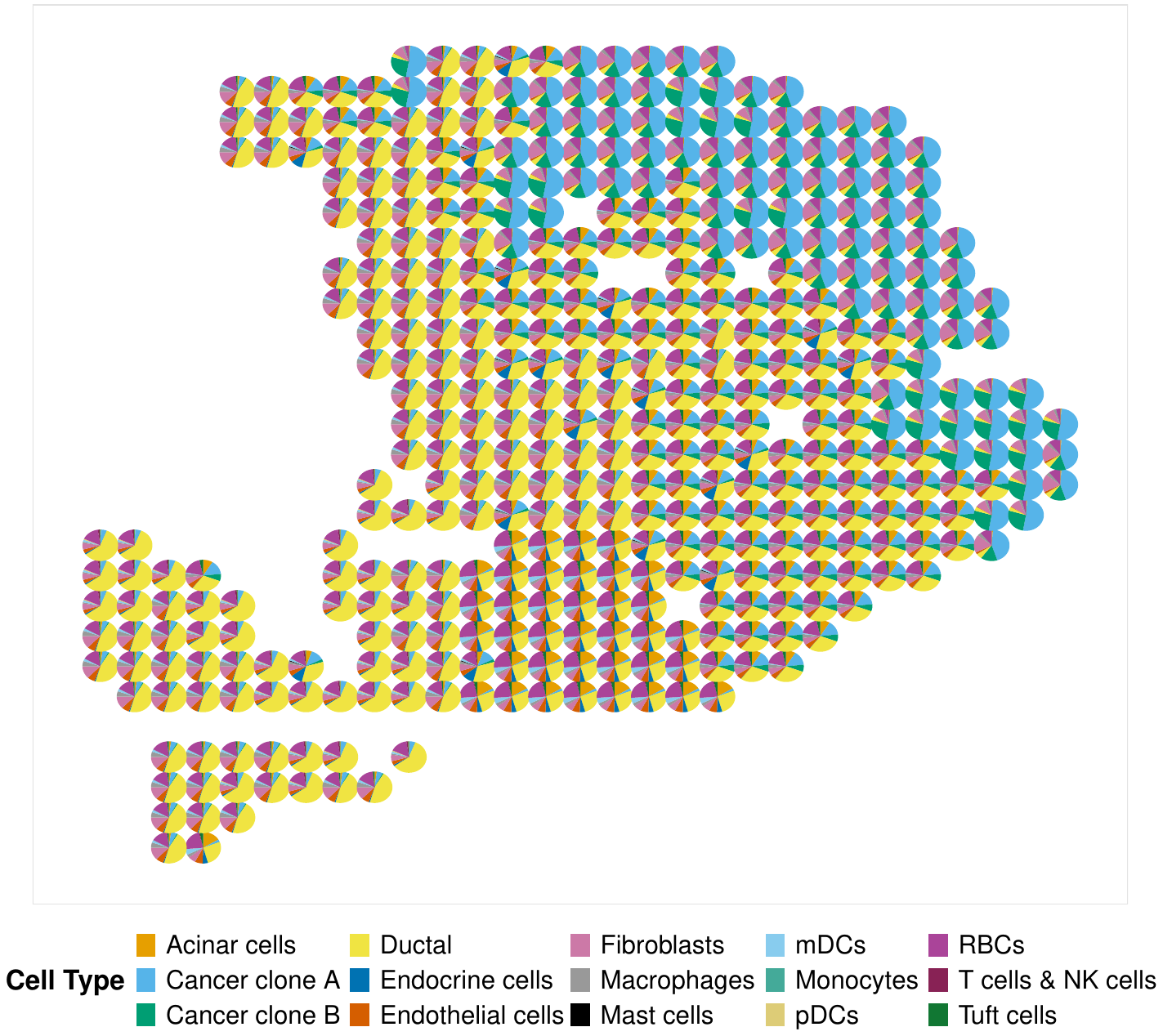}
        \caption{$\widehat\Theta$: BayesSpace}
        \label{fig:sub4}
    \end{subfigure}\hfill
    \begin{subfigure}{0.32\textwidth}
        \centering
         \includegraphics[width=1\textwidth]{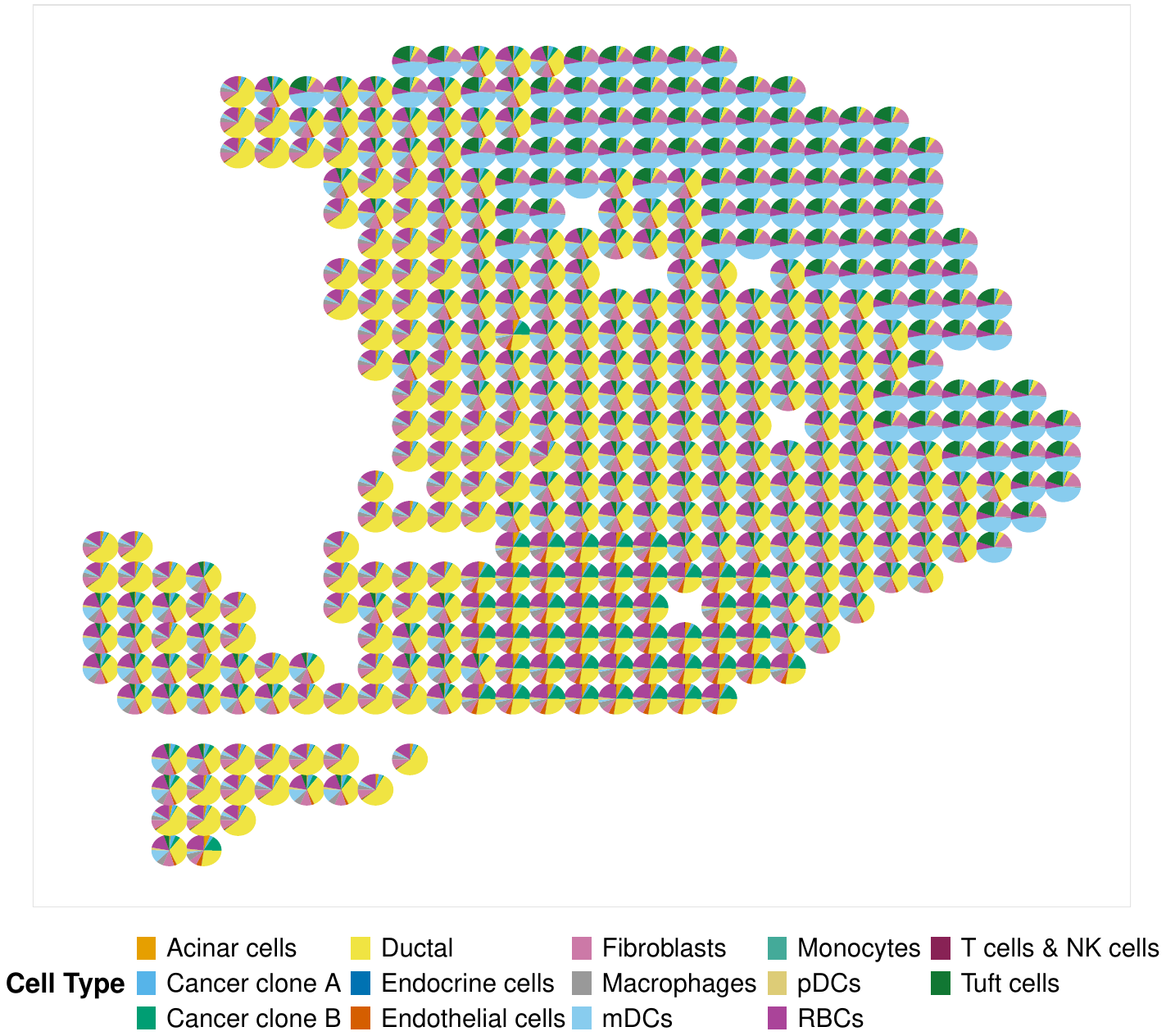}
        \caption{$\widehat\Theta$: Seurat}
        \label{fig:sub6}
    \end{subfigure}
    \vspace{0.2cm} 
    \begin{subfigure}{0.32\textwidth}
        \centering
        \includegraphics[width=1\textwidth]{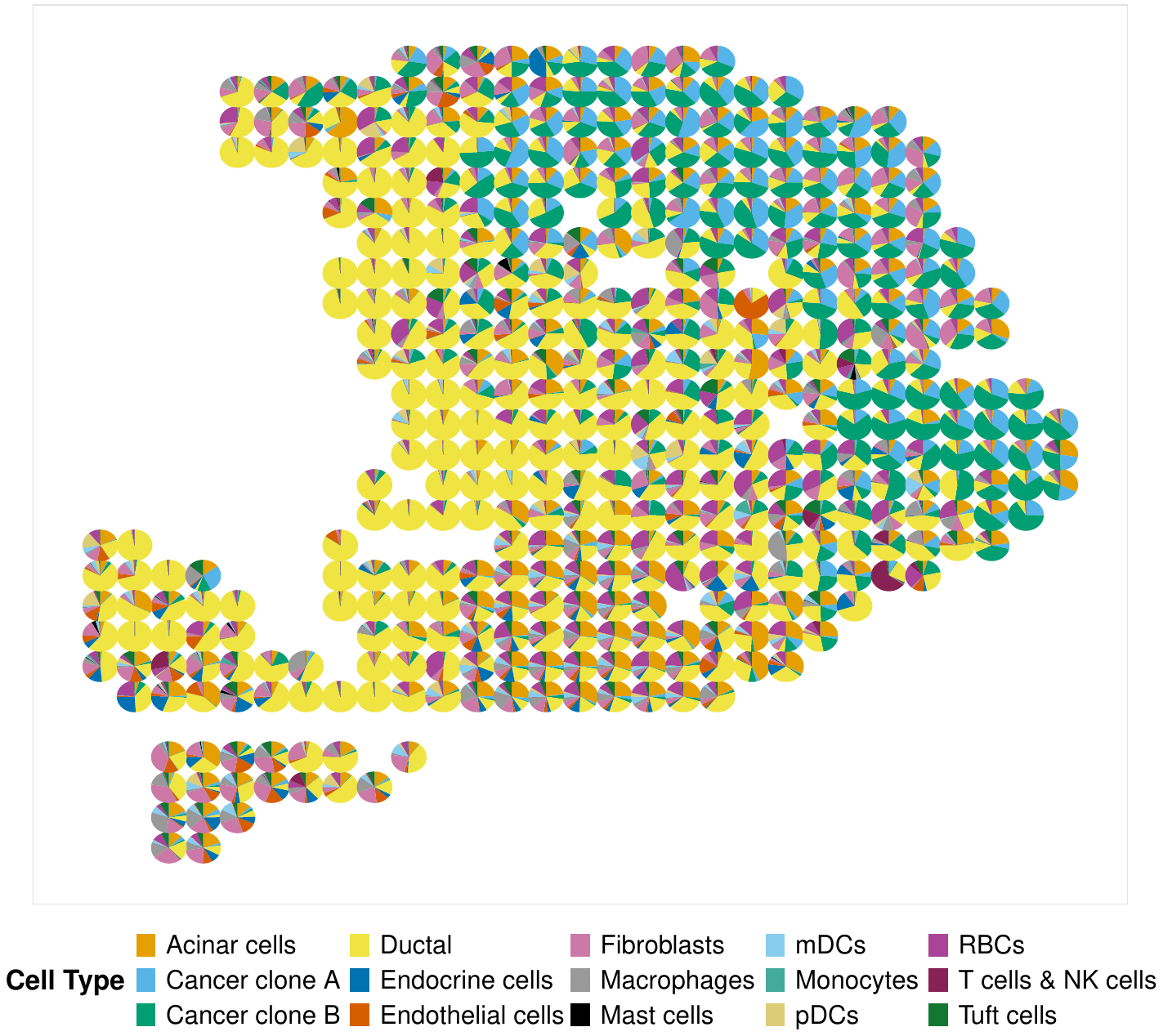}
        \caption{$\widehat{V}$: CARD}
        \label{fig:sub1}
    \end{subfigure}\hfill
    \begin{subfigure}{0.32\textwidth}
        \centering
        \includegraphics[width=1\textwidth]{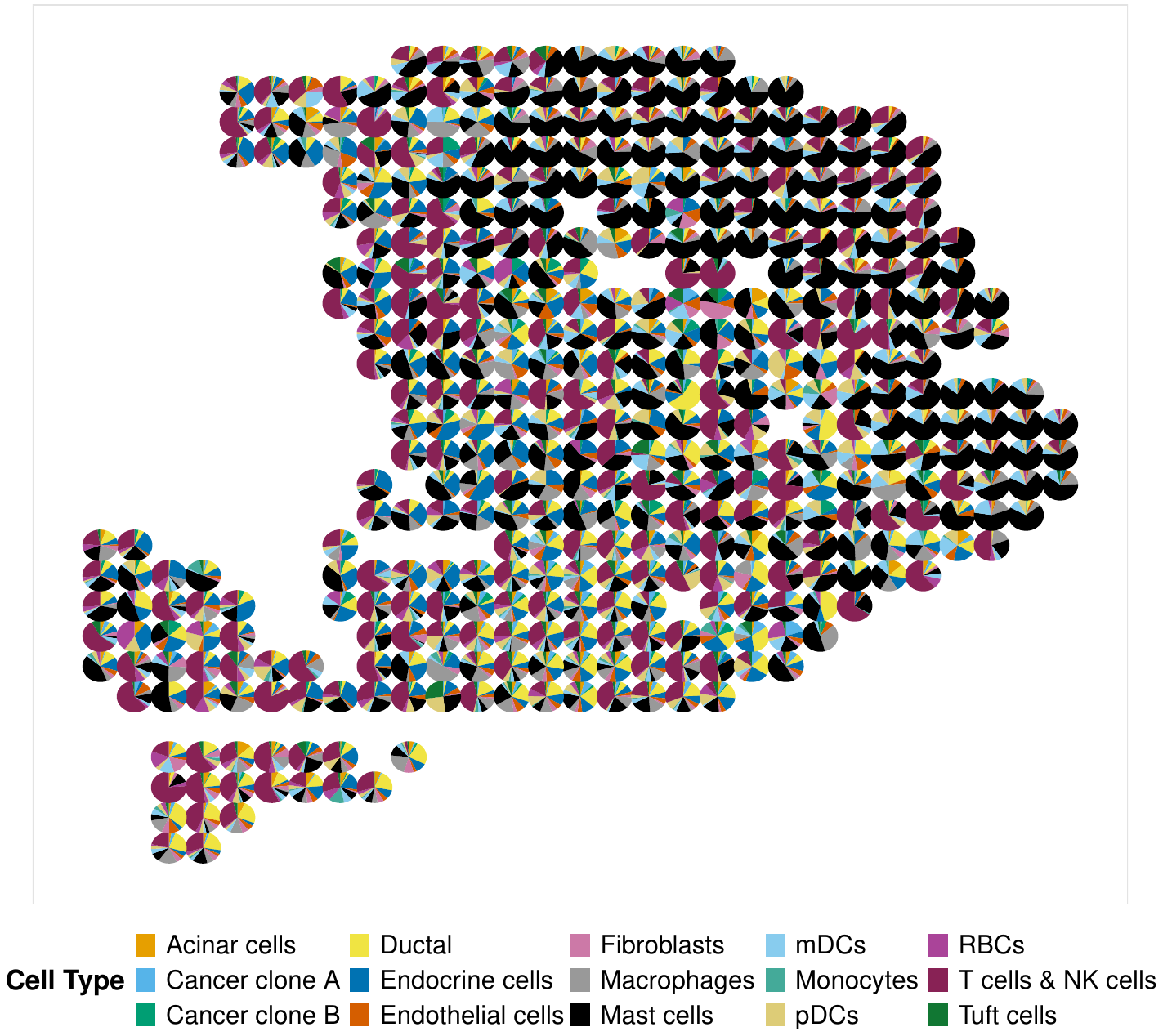}
        \caption{$\widehat{V}$: SPOTlight}
        \label{fig:sub2}
    \end{subfigure}\hfill
    \begin{subfigure}{0.32\textwidth}
        \centering
         \includegraphics[width=1\textwidth]{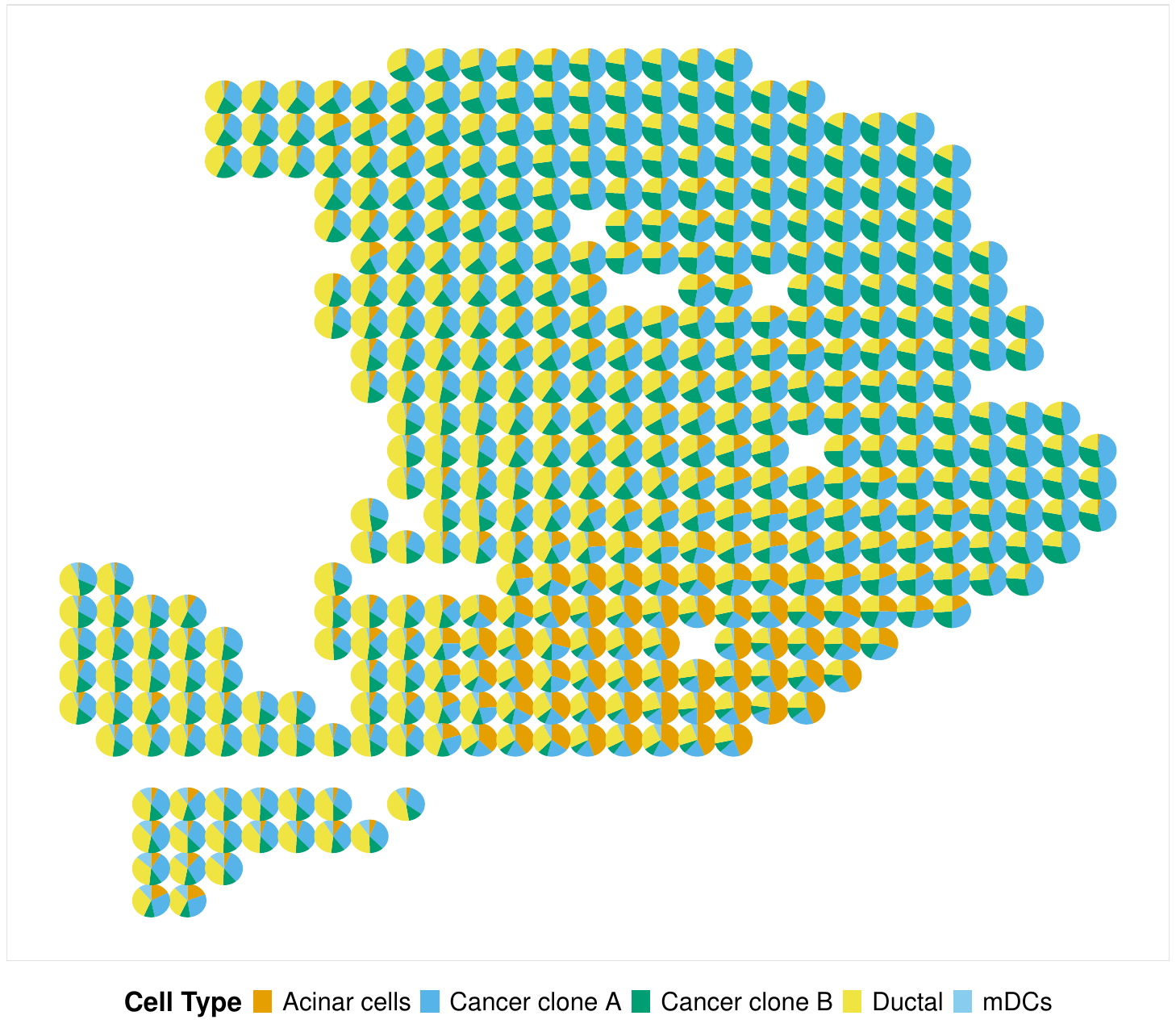} 
        \caption{$\widehat{V}$: IRIS} 
        \label{fig:sub3}
    \end{subfigure}
    \caption{PDAC results showing spatial clusters and their estimated cell-type compositions using different methods: (a) DUET, (b) BayesSpace, (c) Seurat, (d) CARD, (e) SPOTlight, and (f) IRIS. Each pie chart represents the estimated cell-type proportions within a spot, with colors corresponding to different cell types. Note that DUET, BayesSpace, and Seurat provide estimates of the expected cell-type proportions, whereas IRIS, CARD, and SPOTlight provide estimates of the realized cell-type proportions.}
    \label{fig:PDAC_comp_results}
\end{figure}

\vspace{-10pt}
\subsection{Results on PDAC}
We constructed spatial weights as described in Section~\ref{s:implementation} with $k^*$ and $k^{**}$ both set to 5. Applying our method with tuning parameter selection via BIC identified spatially coherent clusters across the PDAC tissue section.



Figure~\ref{fig:PDAC_comp_results} shows the inferred cell-type proportions across the tissue section, which can be directly compared against the discrete histological annotations (Figure~\ref{fig:sub1}). While the histological annotations provide a macroscopic, single-label categorization of the tissue into cancer, ductal epithelium, stroma, and normal pancreatic regions, our deconvolution method reveals the underlying continuous cellular heterogeneity that characterizes the PDAC microenvironment.

We observe compositional shifts that align with the broad histological regions while capturing localized mixing. In the upper right regions annotated histologically as cancer, our method detects the presence of Cancer clones A and B. Also, our results show that these malignant regions are not homogeneous but rather, they are a mixture of fibroblasts, macrophages, and endothelial cells. This co-localization reflects the hallmark desmoplastic reaction of PDAC, where tumor cells are embedded within a dense stromal matrix. Moving toward the center of the tissue which is annotated primarily as stroma, our method captures highly diverse cellular microenvironment. This intermediate zone serves as a transitional bridge, marked by a rich mixture of fibroblasts, ductal cells, and red blood cells, indicating an area of active tissue remodeling and vascularization. On the left margin, regions annotated as ductal epithelium show corresponding ductal cell populations intermixed with structural stroma. Finally, in the lower right regions annotated as normal pancreatic tissue, our method identifies populations of acinar and endocrine cells, consistent with healthy exocrine and endocrine pancreatic function.

Overall, our method successfully recovers a nuanced, spatially coherent map. These results demonstrate our method's ability to capture complex heterogeneity that define pancreatic cancer.

\vspace{-10pt}
\subsection{Additional mouse olfactory bulb SRT data analysis}
In the Supplementary Materials Section 4, we also analyze paired SRT and scRNA-seq datasets derived from the mouse olfactory bulb (MOB), originally generated and published in separate studies. The spatial transcriptomics data were obtained from the publicly available dataset from \citet{staahl2016visualization}, accessible at the Spatial Research repository (\url{https://www.spatialresearch.org}). Speaking broadly, the relative performances of the various methods largely cohere with those seen in the PDAC data analysis. 

\vspace{-10pt}
\section{Discussion}
\label{s:discuss}
\vspace{-5pt}
There are two extensions and generalizations of DUET that we did not address in the article, but are of clear practical interest. First, DUET can be naturally used in settings where larger tissues are analyzed across multiple slides. In such an analysis, we could simply incorporate all spots from all slides into the loss function, then create weights $\gamma_{ij}$ that ``stitch'' the boundaries of various tissue parts. Second, we are presently developing a reference-free variation of DUET. Notably, this requires reformulation since identifiability issues can arise when $B$ is also estimated from the SRT data. 

\setstretch{1}
\bibliography{literature3.bib}

\end{document}